# DELVING INTO YOUTH PERSPECTIVES ON IN-GAME GAMBLING-LIKE ELEMENTS

A Proof-of-Concept Study Utilising Large Language Models for Analysing User-Generated Text Data

# EXECUTIVE SUMMARY

## PROJECT SUMMARY

This report documents the development, test, and application of Large Language Models (LLMs) for automated text analysis, with a specific focus on gambling-like elements in digital games, such as lootboxes. The project aimed not only to analyse user opinions and attitudes towards these mechanics, but also to advance methodological research in text analysis. By employing prompting techniques and iterative prompt refinement processes, the study sought to test and improve the accuracy of LLM-based text analysis. The findings indicate that while LLMs can effectively identify relevant patterns and themes on par with human coders, there are still challenges in handling more complex tasks, underscoring the need for ongoing refinement in methodologies.

## KEY FINDINGS

**Performance Variability Across Tasks:** The LLMs showed strong performance in simpler tasks such as identifying financial engagement and gambling comparisons, approaching human-level accuracy. However, in more complex tasks like aspect-based sentiment analysis, the models' performance was less consistent, highlighting areas for further methodological improvement.

**Prompting for Classification:** This study introduced and tested several prompting techniques that demonstrated an impact on the LLMs' ability to classify complex user-generated content, influencing the effectiveness of the models in various text analysis tasks.

**Application in Large-Scale Analysis**: The LLMs efficiently processed and classified 138,439 user comments, identifying that only 0.81% mentioned lootboxes or similar mechanics. In these comments, about 15% discussed financial engagement with lootboxes, and around 10% drew comparisons to traditional gambling. This demonstrates the LLMs' effectiveness in categorizing large datasets with high accuracy and consistency, even when specific topics are not dominant in the overall discourse.

**Reliability of LLM Outputs**: The study found that when using consistent prompting strategies and models settings, LLMs generally can produce highly reliable and reproducible outputs across repetitive tasks, particularly in simpler classification scenarios. This reliability is crucial for scaling the use of LLMs in large-scale text analysis projects.

**Challenges with Real-World Texts:** The diverse and often informal nature of the user-generated content posed significant challenges for the LLMs. These challenges emphasized the importance of continuing to refine the methods and techniques used in LLM-based text analysis to better handle such variability.

## SIGNIFICANCE AND IMPACT

The methodological advancements achieved through this study significantly enhance the application of LLMs in real-world text analysis. The research provides valuable insights into how these models can be better utilized to analyze complex, user-generated content. These findings are particularly relevant for researchers, as they contribute to more nuanced understandings of online interactions and discourses surrounding gambling-like elements in games. Moreover, the refined methodologies have broader applications, offering tools and techniques that can be adapted for various research areas requiring sophisticated text analysis.

# INTRODUCTION AND OBJECTIVES

## PROBLEM STATEMENT

Within the realm of digital entertainment, gambling-like elements such as loot boxes have become increasingly prevalent across video games (Kim et al., 2023). This is due to a shift in the marketing and monetisation strategy of video game developers. The traditional business model of a one-off purchase of a video game is being replaced by alternative revenue models (Etchells et al.; 2022). A base game is either made available for free or offered for purchase, with additional content expansions typically available for a nominal fee (Etchells et al.; 2022; Kristiansen and Severin 2020). These so-called microtransactions are not a prerequisite for access to a video game, but rather an opportunity to obtain virtual benefits within the game (von Meduna et al., 2020). This principle allows developers to continuously integrate new content into their video games and thus generate a constant and sustainable revenue stream (Etchells et al., 2022).

Loot boxes are a well-known form of microtransaction. They are virtual containers containing a random in-game item (Drummond et al., 2020; Zendle and Cairns 2019). Their rise in popularity has notably reshaped market dynamics and has sparked a wave of engagement from diverse user demographics, especially among younger players. Despite their widespread adoption and integration into modern gaming, these elements introduce complex challenges and potential adverse effects (Kim et al., 2023). Chief among these concerns are risks of fostering addictive behaviours and prompting excessive financial expenditure, with some instances potentially leading to the initiation of real gambling habits, as loot boxes exhibit structural similarities to gambling (Drummond and Sauer 2018; Griffiths 2018; King and Delfabbro 2019). In essence, a loot box is an item in a video game, often purchased with real-world money, whose content is uncertain at the time of purchase and subject to chance (Zendle et al., 2020a). Recent studies have shown that engagement with loot boxes is associated with problem gambling. It has also been shown that problem gambling is more prevalent among players who frequently use loot boxes (Brooks & Clark, 2019; Drummond et al., 2020; González-Cabrera et al., 2022; Hall, Drummond, Sauer, & Ferguson, 2021; Kristiansen & Severin, 2020; Li et al., 2019; Rockloff et al., 2021; von Meduna et al., 2020; Wardle & Zendle, 2021; Zendle, Cairns, Barnett, & McCall, 2020; Zendle et al., 2019, 2020b, Zendle & Cairns, 2018, 2019).

As Mattinen et al. (2023) show, existing research has predominantly employed quantitative methodologies to gauge the addiction potential and economic impacts associated with these gaming features. However, such approaches often fail to delve into the finer nuances of how these elements are perceived, portrayed, discussed and experienced by the youth. There is a significant gap in our understanding of the subjective perspectives and personal evaluations held by young gamers. These insights are crucial, as they relate directly to how these gambling-like elements influence player experiences, their engagement levels, and their willingness to invest financially in such game mechanics.

Recognizing the limitations of current data landscapes (Kim et al. 2023; Mattinen et al., 2023), our research endeavours to bridge this gap through the development and application of advanced methodologies designed for the analysis of a large corpus of text data. This approach not only aims to enhance the depth and accuracy of our understanding of young gamers' interactions with gambling-like elements but also seeks to innovate in the field of text analysis itself. By refining and validating new text analysis tools and techniques, our project will potentially help enable more sophisticated examinations of large-scale data, facilitating richer insights into how digital behaviours are shaped by and reflect the embedded gambling-like mechanics.

Additionally, the discourse surrounding these gambling-like elements in digital games is itself a critical area of investigation. How these features are discussed and portrayed in various media, including online forums, reviews, and social media platforms, plays a significant role in shaping public perception and influencing player behaviour. Analysing the narrative and rhetoric used when describing lootboxes and similar mechanisms offers invaluable insights into societal attitudes and potential misconceptions, or concerns held by the community. By developing tools to analyse the discourse, we aim to uncover the subtleties in how these elements are framed and debated publicly. Such an understanding is crucial for developing informed strategies to address the ethical, psychological, and social challenges posed by the integration of gambling-like elements into video games.

This methodological advancement could significantly contribute to the discourse on responsible game design and potentially influence regulatory measures. By focusing on the development of new and easy to use text analysis methodologies, our study will provide a toolkit for researchers and practitioners to extract meaningful patterns and insights from extensive textual interactions in digital entertainment.

## LARGE LANGUAGE MODELS

### WHAT ARE PRETRAINED LARGE LANGUAGE MODELS?

The advent of Large Language Models (LLMs) like ChatGPT has revolutionized various domains of textual analysis, offering novel methodologies and insights in fields ranging from social sciences to computational linguistics. In the ever-evolving landscape of artificial intelligence, the emergence of LLMs represents a shift in our approach to textual analysis and data interpretation. These advanced models, underpinned by vast linguistic databases and sophisticated algorithms, have ushered in a new era across multiple disciplines. From unearthing nuanced insights in social sciences to refining complex tasks in computational linguistics, LLMs are redefining the boundaries of what is currently achievable with textual data (Demszky et al., 2023).

These models are built using deep learning techniques and typically utilize a specific architecture known as transformers. Deep learning, a subset of machine learning, involves neural networks with many layers, enabling the model to learn complex patterns and representations from large amounts of data. In the context of language models, deep learning helps in processing and understanding text data to perform tasks such as translation, summarization, classification and text generation.

Transformers, introduced in the paper "Attention is All You Need" by Vaswani et al. in 2017, have become the foundation for most state-of-the-art language models. The transformer architecture relies heavily on a mechanism called attention. The attention mechanism allows the model to weigh the importance of different words in a sentence, irrespective of their position, helping the model focus on relevant parts of the input text when generating outputs. Traditional transformers consist of an encoder that processes the input text and a decoder that generates the output text.

However, models like GPT (Generative Pre-trained Transformer) use only the decoder part for generating text.

One of the defining features of large language models is their scale. They contain billions of parameters (the weights in the neural network) that are adjusted during training. The sheer size of these models allows them to capture intricate patterns and nuances in language. The impact of LLMs is particularly noteworthy due to their ability to handle complex language tasks that require not only understanding the literal meaning of texts but also inferring subtleties, sarcasm, and cultural nuances. This has made them an invaluable asset in fields like social and political science, where understanding the intricacies of human communication is crucial.

Another significant aspect of LLMs are their pre-training and fine-tuning methodology. In the pre-training phase, the model is exposed to a vast corpus of textual data, enabling it to learn a wide range of language patterns, structures, and nuances. This training involves unsupervised learning, where the model self-learns by predicting the next word in a sentence. The fine-tuning phase follows, wherein the model is further trained on a specific dataset to tailor its capabilities to particular applications or domains. This combination of transformer architecture, extensive pre-training, and subsequent fine-tuning results in a model that exhibits remarkable linguistic capabilities (Yenduri et al., 2023). LLMs can engage in tasks such as language translation, question-answering, summarization, and text generation with unprecedented effectiveness and accuracy. Moreover, their ability to adapt to different styles, contexts, and subject matters makes them a versatile tool across various domains.

## RECENT DEVELOPMENTS OF LLMS IN SOCIAL SCIENCE RESEARCH

This section delves into the advancements and applications of Large Language Models in various domains of text analysis. By exploring recent studies, we aim to highlight the progress made in aspect and query-based text summarization, multilingual psychological text analysis, political text annotation, sentiment analysis, and the scaling of political ideologies. These investigations underscore the transformative potential of LLMs, in automating complex tasks traditionally performed by human experts, thereby enhancing efficiency and accuracy across multiple fields of text analysis.

Aspect and Query-Based Text Summarization

The study by Yang et al. (2023) represents an advancement in the field of text summarization, particularly in the context of aspect and query-based approaches. Their research focused on evaluating the performance of ChatGPT in creating summaries based on specific aspects or queries within a text. This type of summarization goes beyond traditional methods by focusing on elements or questions within the text, providing a more targeted and relevant summary. Yang et al. (2023) tested ChatGPT across various benchmark datasets that included diverse types of text, such as Reddit posts, news articles, dialogue meetings, and stories. The findings were notable for demonstrating that ChatGPT's performance was on par with traditional fine-tuning methods in terms of Rouge scores, a common metric for evaluating the quality of text summaries. This suggests that ChatGPT can effectively understand and extract key aspects or queries from texts and generate concise, coherent summaries that align closely with human-generated summaries. These findings are particularly relevant in a information-rich world, where the ability to summarize large volumes of text quickly and accurately is increasingly valuable. ChatGPT's proficiency in aspect and query-based text summarization indicates its potential to transform the way information is processed and presented.

Multilingual Psychological Text Analysis

Rathje et al. (2023) explored the capabilities of GPT, the underlying model of ChatGPT, in the realm of automated psychological text analysis, with a specific focus on its application across multiple languages. The study evaluated the performance of GPT-3.5 and GPT-4 in accurately detecting psychological constructs such as sentiment, discrete emotions, and offensiveness across 12 languages, including English, Arabic, Indonesian, Turkish, and several African languages. The findings were remarkable, indicating that GPT significantly outperformed traditional English-language dictionary-based text analysis methods. Moreover, in some instances, GPT demonstrated comparable or superior performance to fine-tuned machine learning models. This research is a testament to the versatility of GPT models, highlighting their potential to offer high accuracy in psychological text analysis across multiple languages. The ease of use of these models, requiring no extensive training data and being accessible through simple prompts, makes them a powerful tool in the arsenal of researchers and practitioners in psychology and behavioural sciences. The study by Rathje et al. (2023) not only underscores the effectiveness of GPT in automated text analysis but also paves the way for more inclusive and cross-linguistic research, particularly with under-researched languages, thereby broadening the scope and reach of psychological studies.

Annotating Political Affiliation

Törnberg (2023) undertook a critical assessment of ChatGPT-4's capabilities in classifying the political affiliation of individuals based on their Twitter posts. This research is particularly significant in the context of political science and social media analytics, where understanding the political leanings and affiliations of social media users is crucial for analysing political discourse and public opinion trends. Törnberg's evaluation revealed that ChatGPT-4 not only achieved a higher accuracy in correctly identifying the political affiliation of Twitter posters but also demonstrated greater reliability compared to human classifiers. This finding is remarkable, as it highlights the model's advanced understanding of nuanced political language and its ability to make inferences based on the context and content of social media posts. The study highlights ChatGPT-4's potential in automating the task of political text analysis, which has traditionally been a time-consuming and labour-intensive process requiring extensive human involvement. The ability of ChatGPT-4 to perform this task with higher accuracy and consistency opens up new possibilities for researchers and practitioners in political analysis, enabling them to conduct more comprehensive and efficient studies on political discourse and public opinion dynamics.

Text Classification, Public Opinion Analysis and substituting Human Experts

Chae and Davidson's (2023) exploration of LLMs in text classification extends significantly into public opinion analysis, especially in political science and sociology. The use of LLMs in stance detection and sentiment analysis is particularly noteworthy. Stance detection involves categorizing documents with labels such as "Support/Oppose" or "Favour/Against," along with a neutral category for ambivalent or irrelevant texts. This method is closely related to sentiment analysis, which classifies the overall tone of a text as "Positive," "Negative," or "Neutral.". Chae and Davidson illustrate this with experiments using GPT-3 (Davinci) for stance detection. They demonstrate that, while the model can generally provide correct outputs, it initially conflates stance with sentiment. However, when provided with an example (one-shot learning), the model more accurately identifies stances rather than just sentiments. This distinction is crucial in analysing political texts, where the stance towards a subject can be more nuanced than a mere positive or negative sentiment. To expand on these findings, it is essential to understand the broader implications for public opinion analysis in the digital age. With the proliferation of social media, platforms like Twitter and Facebook have become vital for shaping and reflecting public opinion. LLMs, with their nuanced understanding of language and sentiment, offer an unprecedented tool for analysing these vast datasets. They can identify trends, shifts in public sentiment, and even predict political outcomes based on the analysis of social media discourse. This capability is not just limited to English but extends to multiple languages, reflecting the global nature of social media and political discourse.

Rytting et al. (2023) investigate the application of GPT-3 as a synthetic coder across several domains within social science research. Their study covers four main datasets: descriptions of political partisanship (Pigeonholing Partisans), policy issues in New York Times headlines, summaries of U.S. Congressional hearings, and reader responses on populism from The Guardian. In each domain, GPT-3 demonstrated comparable performance to human coders, successfully coding texts related to political stereotypes, media and policy analysis, legislative documentation, and populist rhetoric. Notably, few-shot prompts with examples were used to achieve this performance. This research underscores the versatility of GPT-3 in handling diverse and complex classification tasks across different social science disciplines, highlighting its potential to enhance the efficiency and accuracy of text analysis in these fields.

The research conducted by Heseltine and Clemm Von Hohenberg (2023) explored the efficacy of GPT-4 in the realm of political text annotation, a task that involves analysing and categorizing text based on political content, sentiment, and ideology. This study is crucial for understanding the potential of LLMs in automating complex annotation tasks that have traditionally relied on human expertise. Their findings demonstrated that GPT-4 is highly accurate in annotating political texts, especially for shorter formats like tweets. This level of accuracy in shorter texts is significant as it reflects the model's ability to grasp the essence of political messages conveyed in a concise format, a common characteristic of social media communication. The study further introduced a "hybrid" coding approach, where disagreements between multiple GPT-4 runs were judged by a human expert, leading to even higher accuracy levels. The implications of this research are far-reaching for the field of political text analysis. The ability of GPT-4 to accurately annotate political texts suggests that LLMs can be a viable and cost-effective alternative to human experts in certain scenarios. This advancement holds the promise of transforming the landscape of political analysis, making it more efficient and scalable, and paving the way for more sophisticated and nuanced understanding of political discourse.

Sentiment Analysis with LLMs

Zhang et al. (2023) embarked on a comprehensive investigation into the capabilities of Large Language Models in the domain of sentiment analysis. This study is particularly significant in the current landscape of natural language processing (NLP), where understanding and interpreting human sentiments and opinions from text are crucial for various applications ranging from customer feedback analysis to social media monitoring. The researchers evaluated LLMs across a spectrum of sentiment analysis tasks that included both conventional sentiment classification and more intricate challenges like aspect-based sentiment analysis and multifaceted analysis of subjective texts. This extensive evaluation spanned 26 datasets, covering a wide range of sentiment analysis scenarios. The findings of Zhang et al. (2023) revealed a nuanced view of the capabilities of LLMs. While LLMs exhibited satisfactory performance in simpler sentiment analysis tasks, they showed limitations in handling more complex tasks that required a deeper understanding of the text or structured sentiment information. This insight is vital as it underscores the areas where LLMs excel and where they need further development, thus guiding future research efforts in enhancing their sentiment analysis capabilities.

In their exploratory study, Wang et al. (2023) provided a critical evaluation of ChatGPT's (GPT-3) abilities as a sentiment analyser. Wang et al. (2023) conducted their evaluation in various settings, utilizing 18 benchmark datasets and encompassing five representative sentiment analysis tasks. The study compared ChatGPT's performance with that of fine-tuned Bidirectional Encoder Representations from Transformers (BERT) models and state-of-the-art systems in these tasks. A notable aspect of this study was the examination of ChatGPT's zero-shot performance, a scenario where the model is used without any task-specific training. The results highlighted by Wang et al. (2023) displayed ChatGPT's impressive zero-shot performance in sentiment analysis, demonstrating its ability to rival even fine-tuned BERT models. Furthermore, ChatGPT exhibited a remarkable capacity to handle polarity shifts in text, a common challenge in sentiment analysis where the sentiment changes due to negations or subtle language nuances. This preliminary evaluation by Wang et al. (2023) not only attests to the potential of ChatGPT as a versatile sentiment analyser but also sets the stage for further exploration into its capabilities and limitations in various sentiment analysis contexts.

Recent research by Krugmann and Hartmann (2024) also investigates the proficiency of Large Language Models such as GPT-3.5, GPT-4, and Llama 2 in sentiment analysis, particularly within marketing research. Their study benchmarks these LLMs against high-performing transfer learning models, revealing that despite their zero-shot nature, LLMs can sometimes surpass traditional methods in sentiment classification accuracy. The research explores the impact of textual data characteristics and analytical procedures on classification accuracy, noting that factors such as text complexity and data origin significantly influence LLM performance. For instance, lengthy, content-rich words improve classification accuracy, while single-sentence reviews and less structured social media texts reduce it. The study further evaluates the explainability of sentiment classifications generated by LLMs, finding that models like Llama 2 offer advanced human-like reasoning capabilities. This analysis underscores the transformative potential of LLMs in digital marketing.

Scaling Ideologies of Politicians

Wu et al. (2023) conducted a pioneering study using ChatGPT (ChatGPT-3.5) to assess and scale the ideological positions of members of the U.S. Senate. This study represents a significant advance in the application of LLMs to political science, particularly in the measurement of latent constructs such as political ideology. The researchers utilized ChatGPT to perform pairwise liberal-conservative comparisons between senators, thereby generating a new scale of ideological positioning. This approach leveraged the vast knowledge base and inferential capabilities of ChatGPT, allowing for a nuanced and comprehensive understanding of political ideologies. The findings of Wu et al. (2023) were notable for their strong correlation with established liberal-conservative scales like DW-NOMINATE, affirming the reliability and validity of the LLM-based approach. Moreover, their method offered interpretative advantages, such as providing more nuanced placements of senators who vote against their party lines. This study not only demonstrates the potential of LLMs in political analysis but also suggests new avenues for utilizing these models in social science research, particularly in areas where traditional data collection methods are limited or challenging.

Prompting Strategies for LLMs

Recent research has also explored advanced prompting strategies to enhance sentiment analysis using Large Language Models. Wang and Luo (2023) introduced the Role-Playing (RP) prompting strategy, where the LLM is assigned a specific role, such as an expert in sentiment analysis. This approach significantly improved the model's accuracy across various domains by providing a contextual framework that aligns with human expertise. Additionally, Wang and Luo (2023) used the Chain-of-Thought (CoT) prompting strategy. The combined RP-CoT prompting strategy, which merges Role-Playing and Chain-of-Thought approaches, delivered the best performance. They demonstrated improvements in sentiment analysis accuracy across datasets from movie reviews, finance, and shopping domains. This strategy was especially effective in enhancing the model's performance on implicit sentiment classification, addressing the challenges of identifying sentiments not explicitly stated in the text. This highlights the potential of prompting strategies in NLP, potentially contributing to the development of more accurate and reliable text analysis tools.

Kuila and Sarkar (2024) investigate the efficacy of smaller open-source LLMs in predicting entity-specific sentiment in political news articles using zero-shot and few-shot learning strategies. Their study leverages the Chain-of-Thought approach augmented with rationale in few-shot in-context learning, demonstrating that LLMs outperform fine-tuned BERT models in capturing entity-specific sentiment. The evaluation shows that in-context learning significantly enhances model performance, while the self-consistency mechanism improves consistency in sentiment prediction. Despite these promising results, the effectiveness of the CoT prompting method shows some inconsistencies. Overall, their findings highlight the potential of LLMs in entity-centric sentiment analysis within political news and emphasize the importance of appropriate prompting strategies.

Reiss (2023) conducted a study to test the reliability of ChatGPT for text annotation and classification tasks. The study focused on the non-deterministic nature of ChatGPT, which means that (near) identical inputs can lead to different outputs. This variability prompted an investigation into the consistency of ChatGPT's zero-shot capabilities, particularly in differentiating website texts into "news" and "not news". The findings revealed that minor wording alterations in prompts could result in varying outputs. Although pooling outputs from multiple repetitions can enhance reliability, the study underscores the necessity for thorough validation, including comparisons against human-annotated data.

Due to recent advancements in Large Language Models within the broader field of social science research, applying them to analyse user-generated content regarding lootboxes is emerging as a promising approach. Recent advancements demonstrate that LLMs can automate complex tasks traditionally performed by human experts with impressive efficiency and accuracy. This is particularly relevant for discussions about lootboxes, where the specific words and colloquial language used by users play a crucial role in understanding public sentiment and perceptions. LLMs could potentially excel in parsing and interpreting these nuanced expressions, enabling a deeper and more precise analysis of the conversations surrounding lootboxes. Applying these capabilities to the analysis of loot box discussions can provide valuable insights into user opinions and behaviours and public sentiment, potentially leading to more informed decisions and policies in this area.

## PROJECT OBJECTIVES

A key objective of our proof-of-concept study is to develop and test the use of Large Language Models for automated text analysis, focusing on text data with a thematic emphasis on gambling-like elements in digital games like lootboxes and similar mechanics. This methodological research directly engages with real text data, embracing its inherent complexities and ambiguities, to maintain high external validity in our findings.

**Advantages of this approach:**

Utilizing LLMs for text analysis offers several significant advantages over traditional manual coding methods. The simplicity of implementation stands out, as LLMs can process large volumes of text data quickly and efficiently. This capability is particularly beneficial given the unstructured and idiosyncratic nature of user-generated content, which often includes informal language, slang, and emergent gaming terminology. Moreover, the costs associated with using LLMs are considerably lower compared to manual methods, reducing the need for extensive time and resource investment in data analysis. Additionally, LLMs require minimal preparation of text data for analysis, potentially reducing the need for extensive NLP expertise among users. This opens up this form of analysis to a broader range of researchers who may not possess specialized knowledge in natural language processing, thereby expanding the scope and accessibility of research in this area. Furthermore, the multilingual capabilities of LLMs enhance their applicability across different linguistic contexts, enabling the analysis of sentiments and themes in multiple languages without the need for language-specific tuning. This is particularly advantageous in global gaming communities where players may interact and express themselves in a variety of languages, ensuring that insights gained are not limited by linguistic boundaries.

**Influence of Prompts:**

Another key objective of our study is to compare and evaluate different prompting techniques that can significantly influence the behaviour and performance of LLMs (Wei et al. 2022). In the field of computational linguistics, the use of 'prompts' has become critically important, particularly in relation to Large Language Models which often operate on a decoder-only architecture. These models leverage statistical methods to interpret and produce linguistic patterns. Essentially, a 'prompt' acts as the initial input that informs these models, influencing the generation of subsequent text tokens. For example, in applications like OpenAI's ChatGPT, a user's question serves as a prompt that directs the response of the model. The detailed construction and nuances of a prompt are crucial in determining how the model will respond. This concept is somewhat analogous to human thinking, where the way a question is posed can lead to different reactions and answers.

Through carefully crafted prompts and techniques, the models can be guided to perform deeper and more contextually relevant analyses. Known as Prompt Engineering/Techniques/Tuning, this approach not only allows us to precisely steer and optimize the outcomes of automated text analysis but also enables a systematic comparison of various prompting strategies. By evaluating the effectiveness of different prompts, we aim to enhance the quality and relevance of the insights obtained, and identify which techniques are most effective in different contexts. Prompt engineering thus plays a pivotal role in refining how LLMs interact with complex datasets, ensuring that the analysis is finely tuned to align with specific research questions and contexts.

**Development of Workflows and Tools:**

An integral part of the project is the development and refinement of workflows and tools specifically tailored to the work with LLMs. These developments are designed to facilitate interactions with LLMs and increase the efficiency of our research efforts. The new tools and workflows should not only benefit this project but also support future research endeavours in similar areas. By creating robust, easy to adapt tools and workflows, we aim to enhance the accessibility and effectiveness of LLM technology for researchers across various fields and contexts.

By integrating advanced LLMs and developing specialized analysis tools, we ultimately aim to deepen our understanding of the impacts of gambling-like elements in digital games. At the same time, we are committed to creating sustainable research tools that can be applied beyond the current project, paving the way for innovative approaches in the field of digital content analysis. This project not only advances our understanding of specific content areas but also significantly contributes to the broader field of text analysis by demonstrating how and if advanced computational techniques can be effectively applied to achieve insightful and externally valid results. Specifically, it addresses the challenge of interpreting the distinct expressions and languages of various subcommunities within the gaming world. These subcommunities often develop their own vernaculars and modes of communication, which can include jargon, slang, and symbolisms unique to their group dynamics and gaming experiences. Exploring this possibility, we envision that the sophisticated text analysis enabled by LLMs could reveal subtle nuances in communication that are often overlooked. Understanding these nuances is crucial for accurately analysing and interpreting user-generated content in digital games, providing richer, more context-sensitive insights.

## TARGET DIMENSIONS: SENTIMENT AND THEMATIC ANALYSIS OF GAMBLING-LIKE ELEMENTS

To develop text classification tools and to validate them, it is essential to define clear target dimensions. In our research, we have strategically chosen to employ both aspect-based sentiment analysis (ABSA) and thematic coding. This approach is strategically designed to address typical questions surrounding these elements while tackling various classic text analysis tasks that also differ in task difficulty.

**Aspect-Based Sentiment Analysis (ABSA):**

ABSA is specifically deployed to dissect and categorize the emotional responses players express towards features such as lootboxes, card packs, and similar mechanisms. This method hones in on how sentiments are directly tied to these elements, providing precise insights into player

reactions—whether they view these elements positively, negatively, or neutrally. ABSA is adept at filtering complex and mixed emotional data, focusing solely on the sentiments related to specific gaming components. This not only provides clarity but also tackles one of the more challenging aspects of text analysis—accurately discerning specific sentiments within texts where multiple subjects and sentiments might be interwoven. It is particularly effective in distinguishing clear emotional tones in discussions about contentious features. Another term for this type of task is Entity-Centric Sentiment (Kuila and Sarkar, 2024), as the focus is not on the sentiment of the entire text or its parts, but specifically on a particular entity, in this case, lootbox mechanisms. Nevertheless, we will continue to use the more common term ABSA.

**Thematic Coding:**

Thematic coding is applied to uncover broader themes regarding how gambling-like elements impact the gaming experience, player investment, and perceptions of these elements as gambling. This aspect is excellent for managing the inherent complexity of text data, especially when discussions are nuanced and spread across a vast corpus. Thematic coding excels in organizing these diverse narratives into digestible and analysable segments, which is crucial for addressing broader research questions about the integration and impact of these elements in games.

We have used the following three topics as exemplary categories for thematic coding:

**Gaming Experience:** This category involves determining whether gambling-like elements influence core aspects of the gaming experience, including gameplay, progression, balance, fairness, and immersion. It is about understanding the functional impacts rather than emotional reactions, providing a more objective measurement of how these features integrate into and affect the game mechanics.

**Financial Engagement**: Here, the focus is on whether users discuss spending real money on these elements. Thematic coding in this context aims to capture discussions around financial transactions, distinguishing between mere usage and actual expenditure on gambling-like elements.

**Gambling Comparisons**: This topic looks for explicit, direct comparisons of in-game features with traditional gambling. This thematic coding is critical for identifying how players and the public perceive resp. discuss these mechanisms in relation to established gambling activities, which may also have significant implications for legal and ethical considerations.

By categorizing the first dimension as "Aspect-Based Sentiment Analysis," we emphasize the granularity with which sentiments are analysed in relation to specific game features. The other dimensions, grouped under "Thematic Coding," highlight the theme-oriented approach that looks beyond sentiment to explore broader topics such as impact on gameplay, financial involvement, and legal perceptions.

The first two dimensions, aspect-based sentiment analysis and gaming experience, present a higher level of difficulty due to their intricate nature. ABSA requires a detailed understanding of player sentiments and the ability to distinguish between nuanced emotional responses towards specific features. This involves not only recognizing positive, negative, or neutral sentiments but also understanding the context and reasons behind these expressions and the specific target/entitiy of those sentiments. Additionally, the presence of multiple categories makes the classification task more challenging, as demonstrated in other studies (e.g., Wang and Luo 2023; Krugmann and Hartmann 2024).

Similarly, analysing the gaming experience dimension demands a deep comprehension of how gambling-like elements influence various aspects of gameplay. This includes assessing changes in aspects like game quality, game balance, fairness, and overall immersion. Such analysis requires a meticulous approach to ensure that subtle yet significant impacts on the gaming experience are accurately identified and categorized.

On the other hand, the dimensions of financial engagement and gambling comparisons, while still complex, tend to be more straightforward. Financial engagement focuses on identifying discussions around monetary transactions related to gambling-like elements. This involves tracking mentions of spending, which is generally more direct and less ambiguous compared to emotional sentiments or nuanced gameplay impacts.

Gambling comparisons involve detecting explicit comparisons between in-game features and traditional gambling. This task is more focused on identifying specific language and phrases used by players to draw parallels between the two, making it somewhat easier to classify compared to the more abstract and multifaceted dimensions of sentiment and gaming experience.

With these target dimensions of varying difficulty, we aim to investigate the classification capabilities of LLMs. By differentiating between the nuanced emotional analyses provided by ABSA and the broader contextual insights offered by thematic coding, we can construct a comprehensive picture of how gambling-like elements are integrated, perceived, discussed, and depicted within the gaming community.

# METHODOLOGICAL APPROACH

## RESEARCH DESIGN

### ITERATIVE PROMPT DEVELOPMENT

The overarching objective of this project is to develop optimized prompts for a large language model to effectively analyse user-generated texts, specifically focusing on identifying opinions, sentiments, and evaluations related to lootboxes in video games. To achieve this, we employ an iterative prompt development approach, a systematic methodology that refines and enhances the model's prompts through multiple cycles of evaluation and adjustment. By engaging in this iterative process, we aim to shape prompts that elicit increasingly accurate, nuanced, and contextually appropriate responses from the language model when tasked with analysing texts about lootboxes.

Through this approach, we strive to develop prompts that enable the language model to capture the nuances and complexities inherent in user-generated content, ultimately providing valuable insights into the perspectives and attitudes surrounding the controversial topic of lootboxes in video games. This iterative prompt development process aims to facilitate robust quantitative analysis of large textual datasets focused on lootboxes through optimized prompts.

Iterative prompt development is a crucial process when working with large language models to optimize their outputs. It involves a cyclical refinement of the input prompt through multiple iterations, with the aim of eliciting more relevant, coherent, and comprehensive responses from the model. This systematic approach can be broken down into the following key stages:

1. Initial Prompt Formulation: The process begins with the drafting of an initial prompt that outlines the core subject matter and desired task. This first iteration serves as a baseline for subsequent refinements.

2. Response Evaluation: The language model's output in response to the initial prompt is carefully analysed and evaluated. Criteria for assessment include relevance to the intended topic, coherence and comprehensibility, and adequate depth and detail. This evaluation is inherently a qualitative process, requiring human judgement and interpretation. The qualitative assessment involves closely examining the content, structure, and language use in the model's response. Evaluators must critically analyse whether the key points and information are properly covered, if the logical flow is coherent, and if the response format and phrasing are appropriate for the intended purpose. This nuanced inspection goes beyond mere keyword matching or quantitative metrics.

3. Prompt Refinement: Based on the evaluation, the prompt is iteratively refined and enhanced. This may involve adding or modifying instructions, constraints, or other elements to guide the model's response more effectively.

4. Iterative Cycles: Steps 2 and 3 are repeated in a cyclical manner, with each new iteration building upon the insights and adjustments from the previous round. This iterative process can continue until the model's outputs meet the desired standards of quality and suitability for the intended use case.

Once a well-tested and refined prompt has been developed, it may be adapted and applied to analogous tasks or domains, further leveraging the invested effort and knowledge gained through the iterative development process.

The iterative prompt development approach requires patience, diligent evaluation of intermediate results, and a willingness to engage in multiple cycles of refinement. By incorporating this qualitative dimension into iterative prompt development, a more complete understanding of the model's capabilities and limitations can be gained. This holistic perspective guides meaningful refinements to shape the model's outputs towards the desired level of quality, nuance, and effectiveness. However, this systematic process has proven instrumental in eliciting significantly improved outputs from large language models, ultimately enhancing their utility and practical value across a wide range of applications.

## PROMPTING TECHNIQUES FOR LARGE LANGUAGE MODELS

As large language models continue to advance, effective prompting strategies have emerged as a critical factor in harnessing their full potential. Among the various prompting techniques, zero-shot prompting, chain-of-thought prompting, and collaborative reasoning prompting have garnered attention due to their potential to elicit more coherent and reasoned outputs from these models. Therefore, in our study, we investigate classification using these prompting techniques and analyse how the results differ across the techniques.

**Zero-Shot Prompting:** Zero-shot prompting refers to the approach of providing a language model with a prompt and task instructions without any explicit examples or demonstrations. The model is expected to draw upon its vast pre-trained knowledge to understand the prompt and generate relevant responses. This technique is particularly useful when specific examples are unavailable or when the goal is to evaluate the model's general reasoning and language generation capabilities.

While zero-shot prompting may seem challenging, careful prompt engineering can guide the model towards the desired output. This often involves providing clear and precise instructions, leveraging the model's understanding of natural language, and incorporating relevant context or background information within the prompt itself.

**Chain-of-Thought Prompting:** Chain-of-thought prompting is a more recent technique that aims to elicit step-by-step reasoning from language models, particularly for tasks that require multi-step problem-solving or complex analytical processes. In this approach, the model is prompted to generate not only the final answer but also the intermediate steps and thought processes involved in arriving at that answer.

This prompting strategy is inspired by the observation that humans often engage in a chain of reasoning, breaking down complex problems into smaller steps and explicating their thought process along the way. By encouraging language models to follow a similar chain of thought, the resulting outputs become more transparent, interpretable, and potentially more reliable.

Implementing chain-of-thought prompting involves crafting prompts that explicitly instruct the model to "think aloud" and articulate its step-by-step reasoning. This can be achieved by providing examples or demonstrations of the desired thought process or by incorporating specific instructions within the prompt itself.

**Collaborative Reasoning Prompting (Three of Thought[1]):** Collaborative reasoning prompting is a technique that leverages the collective intelligence of multiple agents or perspectives to tackle complex problems. In this approach, different experts (or simulated expert agents) contribute sequentially to the task, each building upon the previous analysis while sharing their confidence levels regarding the correctness of their assertions. If an expert realizes their approach is flawed, they can leave the process, allowing the remaining experts to refine the solution collaboratively.

This method mimics real-world collaborative problem-solving, where multiple individuals contribute their unique insights and iteratively refine their thoughts based on peer feedback. The final solution is either a consensus among the experts or the best guess based on the cumulative analysis.

---

[1] The name "Three-of-Thought" is derived from our approach of simulating *three* experts within a single LLM call.

Implementing collaborative reasoning prompting involves structuring prompts to guide the sequential and collaborative contribution of different experts, often incorporating mechanisms for confidence rating and peer review to enhance the robustness and reliability of the generated solutions.

**Comparison and Applications:** Zero-shot prompting, chain-of-thought prompting, and collaborative reasoning prompting offer distinct advantages and applications. Zero-shot prompting allows for evaluating a model's general capabilities and versatility, while chain-of-thought prompting provides insights into the model's reasoning process and can enhance trust and interpretability, particularly for complex tasks, as it encourages the model to "think" through problems step-by-step, revealing its logical progression. Collaborative reasoning prompting, on the other hand, harnesses the power of collective intelligence, making it ideal for tasks that benefit from diverse perspectives and iterative refinement.

It is important to note that the effectiveness of these prompting techniques is heavily dependent on the quality of the prompts themselves. Careful prompt engineering, leveraging techniques such as iterative refinement and qualitative evaluation, is crucial for optimizing the prompts and eliciting the desired outputs from large language models.

## MEASURES FOR THE EVALUATION

In our study, we decided to use Cohen's Kappa as an evaluation criterion. The primary reason for this decision lies in the inherent ambivalence of user texts, making it impossible to establish a clear ground truth against which the classification of LLMs could be assessed. User texts are often ambiguous and open to interpretation, meaning there are no definitive right or wrong answers. In such cases, evaluating the performance of LLMs against a predetermined ground truth would be inappropriate, as this ground truth could itself be subjective and contested. The ambivalence regarding the target dimensions is also evident from the disagreement among human coders. This disagreement highlights the subjective nature of interpreting user texts and reinforces the need for evaluation criteria that account for such variability.

Thus, we opted to use Cohen's Kappa, which quantifies the agreement between human coders and the LLM. By measuring the LLMs' agreement with human coders, we can obtain a more objective and fair evaluation criterion. Cohen's Kappa accounts for the possibility of chance agreement and thus provides a more robust metric for assessing consistency between different coders, whether human or machine.

Additionally, we also use Cohen's Kappa to evaluate the agreement between different prompt techniques. This is particularly important as it allows us to compare the consistency and reliability of various techniques and analyse how well they handle the ambivalence of user texts.

Cohen's Kappa values can be interpreted using several benchmarks. While there is no absolute consensus, a widely cited scale proposed by Landis and Koch (1977) is commonly used:

| | | |
|---|---|---|
| **< 0**: | No agreement (worse than chance) | |
| **0.00 – 0.20**: | Slight agreement | |
| **0.21 – 0.40**: | Fair agreement | |
| **0.41 – 0.60**: | Moderate agreement | |
| **0.61 – 0.80**: | Substantial agreement | |
| **0.81 – 1.00**: | Almost perfect agreement | |

Furthermore, alongside Cohen's Kappa, we also consider the percentage agreement of classifications. This metric provides a straightforward measure of the proportion of classifications on which different coders agree, offering additional insight into the performance of the LLMs and the effectiveness of the prompt techniques.

In summary, Cohen's Kappa is the most suitable evaluation criterion in our context, as it captures the agreement between evaluations while considering the inherent ambivalence of user texts. This approach allows for a fairer and more accurate assessment of the performance of LLMs and different prompt techniques.

## DATA SOURCES

Our methodological approach aims to develop prompts that enable LLMs (Large Language Models) to capture and interpret a nuanced understanding of user opinions on loot box mechanisms in games on real world online text data. Our research project focuses on user perceptions of lootboxes and similar mechanics in video games, a topic of growing importance in modern gaming culture. To establish a broad empirical basis for our prompt development for LLMs and to conduct a thorough analysis, we chose user reviews on Steam, one of the largest gaming platforms, as our primary data source. Steam provides a rich and relatively accessible source of user feedback, which, due to its diversity and volume, offers a suitable foundation for our tool and prompt development goals.

The selection of games was based on the list of the most played games on 'SteamCharts' (https://steamcharts.com/). In the last week of November 2023, we manually identified the top 10 games on this list that include loot box mechanisms. These games were selected because their popularity and presence in the gaming community reflect a wide range of user opinions and experiences. The high user activity of these games ensures not only a large number of reviews but also a diversity of perspectives, gaming communities, game mechanics, and group language, all of which are essential for our research objectives.

Using a Python script and the 'steamreviews' library, we collected the latest user reviews of these games, focusing on data from the last 30 days. This time frame ensured that we captured a snapshot of current opinions and impressions from players, which is crucial for assessing the contemporary perception of lootboxes while still providing a sufficiently large data base. These data include not only user comments on the games but also metadata on comment ratings and the associated user accounts. However, no sensitive data are included.

Through this approach, we were able to create a comprehensive and diverse dataset of user reviews for the most popular games with loot box mechanisms. This diversity of data, both in terms of the games and user opinions, forms the foundation for the development and study of our text analysis tools.

This collected user text data was reduced in the next step so that it could be assessed by our human coders and the results of the LLM classifications could be evaluated in turn for the prompt development.

We have proceeded as follows:

The first major step in the filtering process involves filtering the reviews based on language, retaining only reviews written in English and German. This ensures that the analysis is performed on reviews that can be processed effectively given the language proficiencies of our human coders.

Next, we identified specific keywords related to lootboxes and similar gaming mechanisms. Reviews containing any of these keywords were marked for further examination. This step ensured that our dataset was relevant to our research focus on loot box perceptions.

To refine this selection further, we utilized a more sophisticated approach involving Large Language Models from OpenAI. We applied a zero-shot classification technique, allowing the LLM to evaluate each review without prior training on our specific dataset. The LLM analysed each review based on a carefully designed prompt template, which guided the LLM in identifying references to lootboxes and related concepts. This prompt instructed the LLM to determine if a review mentioned lootboxes, card packs, loot crates, Gacha, Mystery Boxes, or similar terms. Following this detailed filtering process, we retained only the reviews that specifically mentioned lootboxes and related concepts.

In a final step, we created a training set, where the sampling of the remaining user texts was reduced to 300. Stratified random sampling was used to ensure that both the games and the languages (English and German) were evenly represented in the comments. It is important to note that the objective of this sampling was not to create a random selection with representativeness of the overall population of texts, but rather to generate a set of highly diverse texts. This diversity was crucial for the development of effective prompts, as it provided a broad range of linguistic variations and contexts to better refine our instructions for the model and prompt engineering techniques.

This filtering was necessary to ensure that only user texts relevant to lootboxes were included, a variety of different games and loot box mechanics were present, and to allowing our human coders to make adequate assessments without being overwhelmed by the volume of data.

For the final analysis of model performance, a validation sample was subsequently drawn, which, aside from the language and keyword selection, was chosen randomly.

## DESCRIPTION OF OUR APPROACH

Our approach involved several coordinated steps to optimize the effectiveness of prompt development and the performance of the LLM models.

In the initial phases, we focused on the output and adherence to the prompt instructions. We used a few example comments to test and refine the first prompt iterations. In doing so, we aimed to ensure that both the instructions of the prompting techniques were followed, and the output format of the classification was adhered to. This early stage allowed us to understand how the model responded to various prompts and provided a foundation for further development.

As a result of this step, we developed modular prompt templates designed to reliably generate different prompting techniques. These templates were created to be flexible and adaptable, incorporating various requirements and contextual information to guide the model effectively.

In parallel, we developed methods for extracting outputs from the LLM models. These extraction methods helped us systematically retrieve and analyse the results, enabling detailed evaluation and refinement of the prompts.

After these first steps, the whole training data were coded by two human coders. These coders are well-versed in the games, their mechanics, and the language used within the gaming communities, making them highly suitable for the coding tasks. This manual coding served as a reference to assess the performance of the LLM models and the accuracy of the automated analyses. The human-coded data provided a benchmark against which we could measure the model's output.

After developing and testing the initial prompts and methods, we applied the LLMs to the entire dataset. A manual review of the non-matching results was conducted to identify and analyse inconsistencies. This step was crucial in understanding the areas where the model's performance diverged from the human-coded data. This involved carefully reading through the LLM responses, reconstructing their logic, and assessing how well they adhered to the given instructions. By dissecting the reasoning process of the LLMs, we were able to pinpoint specific issues and areas for improvement in the prompt design.

Based on insights from the labour-intensive review, we manually adjusted the prompts. These adjustments aimed to improve the accuracy and relevance of the model's responses. The iterative process of prompt development and adjustment continued until no further improvements in results were observed. This saturation point, achieved after three iterations, indicated that the prompts had reached their optimal performance, excluding minor intermediate steps, tests and adjustments.

## LARGE LANGUAGE MODELS USED IN THE STUDY

For our study, we utilized two prominent large language models, representing both the commercial and open-source domains. This approach allowed us to compare the performance and capabilities of widely recognized commercial models with leading open-source alternatives. Both models were operated with a temperature setting of 0.0, which configures them to generate the most probable next token at each step. This setting aims to maximize consistency and minimize randomness in their outputs. While this does not guarantee fully deterministic responses, it achieves the closest approximation possible.[2]

---

[2] As part of our study, we also examined the performance of models with a temperature of 0.3 and 0.7 and were able to determine a largely identical performance with regard to our

### CHATGPT-4 TURBO (LLM A)

ChatGPT-4 Turbo, developed by OpenAI, is an advanced language model that excels in generating human-like text based on input prompts. It is designed to handle complex conversational tasks and detailed reasoning, making it suitable for various natural language processing applications. Built on the GPT (Generative Pre-trained Transformer) architecture, this model has an estimated number of parameters that may exceed 1.5 trillion, although the exact number is not publicly disclosed. OpenAI has implemented a training method called "Constitutional AI" to produce more ethical and safer outputs. The model's knowledge cut-off is typically April 2023, though this may vary depending on the specific version. OpenAI GPT-4 is rumoured to be based on the Mixture of Experts (MoE) architecture. It consists of multiple (8-16) models, each with 220 billion parameters, linked in the MoE framework (https://lifearchitect.ai/gpt-4/). This architecture is a type of ensemble learning that combines different models, called "experts," to make decisions. A gating network determines the weight of each expert's output based on the input, allowing different experts to specialize in different parts of the input space. The model operates under a proprietary license and is only available via API or web interface.

### META-LLAMA-3-70B-INSTRUCT (LLM B)

Meta-LLaMA-3-70B-Instruct, developed by Meta, is a large language model with 70 billion parameters. It is part of the LLaMA (Large Language Model Meta AI) series, designed for high-performance natural language understanding and generation. It is specifically optimized for following detailed instructions and providing accurate responses. This model is built to handle complex tasks requiring nuanced understanding and precise text generation. Llama 3 was pretrained on a vast dataset exceeding 15 trillion tokens from publicly accessible sources. For fine-tuning, it utilized publicly available instruction datasets alongside more than 10 million examples that were human-annotated. The Knowledge cut-off is December 2023. Notably, Llama 3 is open source, making it accessible for research and development by the wider community. (Meta. (2024). Llama 3 Model Card. Retrieved from https://github.com/meta-llama/llama3/blob/main/MODEL_CARD.md)

---

classification tasks. However, this does not mean that the models exhibit identical reliability, as only the validity of the classifications in aggregate was examined (Reiss, 2023).

## RESULTS

For the results, we aim to compare the performance of human coders with that of large language models using different prompting techniques for four different tasks with variable difficulty (see section Target Dimensions). The LLMs used in this study are ChatGPT-4 Turbo (referred to as LLM A) and Meta-LLaMA-3-70B-Instruct (referred to as LLM B). We present the results of the final iteration of the prompts for the LLMs. The evaluation employs two metrics: Cohen's Kappa, which adjusts for chance agreement, and observed agreement, which measures the raw proportion of agreement between coders. This approach helps to understand the alignment between human and machine classifications. In the context of evaluating the alignment between human coders and machine-generated classifications, employing a reliability measure such as Cohen's Kappa ostensibly appears to assess the reliability of classifications. However, it fundamentally serves as a tool for estimating validity in this scenario, specifically the construct validity that pertains to how well the classifications from the LLMs align with those from human coders, rather than merely assessing the reliability of the machine classifications themselves.

The classifications were performed by two human coders and LLMs using different prompting strategies: Chain-of-Thought (CoT), Three-of-Thought (shorthand for: Collaborative Reasoning) (ToT), and Zero-Shot (ZS). We visualized the evaluations using heatmaps and analysed them with Cohen's Kappa and observed agreement metrics.

To provide the most reliable comparison, we present the results of the training data, consisting of 300 cases, as these are more dependable. The results for the test data, which are based on only 60 texts, generally point in the same direction regarding task performance but are less reliable due to the smaller sample size.[3] These test data results are illustrated in the appendix.

In the final step, we applied our best performing text classification to the entire dataset of user texts. This approach allows us to test the scalability and robustness of our models on a larger scale. Additionally, it provides a descriptive insight into the online conversation surrounding lootboxes.

### RESULTS FOR ASPECT BASED SENTIMENT ANALYSIS

#### HUMAN CODERS

The analysis between human coders, as measured by Cohen's Kappa, showed a substantial agreement with a value of 0.76. This level of consistency is further confirmed by an observed agreement of 0.86, indicating robust reliability in human sentiment classification. These metrics show that human coders are quite consistent, but not perfect, reflecting the inherent ambiguity in aspect-based sentiment analysis for our data (see Figure 1). Therefore, it is also to be expected that the agreement between the LLM prompts and the human coders will be lower.

---

[3] In our estimation, the iterative prompt development process used in this study should be understood as a form of fine-tuning rather than overfitting. This process involved providing general instructions and refining prompts to better align with task requirements, without resorting to case-specific classifications examples. This approach ensures that the LLMs' performance improvements are robust and applicable to a wider range of similar tasks, rather than being tailored to the specific instances seen during training.

### LLM A

When comparing the agreement between human coders and LLM A, Cohen's Kappa values indicated only moderate agreement. Specifically, Human 1 and the LLM A using CoT prompting showed a Kappa value of 0.55, while the agreements with ToT and ZS prompting methods were 0.56 and 0.54, respectively. Human 2 exhibited slightly lower Kappa values with the LLM A: 0.49 for CoT, and 0.48 for both ToT and 0.47 for ZS. These moderate Kappa values are supported by observed agreement values ranging from 0.69 to 0.75, suggesting that while the prompts of LLM A are fairly consistent with human coders, they do not achieve human-level agreement for the ABSA.

### LLM B

Similarly, LLM B showed moderate agreement with human coders. These results mirror the findings for LLM A, only slightly lower, indicating that LLM B also exhibits reasonable consistency but falls short of human-level agreement.

### LLM A VS LLM B

The comparison between the two LLMs shows high internal consistency within each LLM set, with observed agreement values ranging from 0.84 to 0.89 and Cohen's Kappa values between 0.7 and 0.8, indicating substantial agreement between different prompting techniques. However, the comparison between LLM A and LLM B shows slightly lower agreement, with observed agreement values from 0.79 to 0.83 and Cohen's Kappa values between 0.6 and 0.64, indicating some variability in sentiment classification between different models.

This suggests that different prompting techniques yield highly consistent sentiment classifications among LLMs, but also highlights that both the choice of the specific LLM and the prompting technique make a difference since they are not perfectly in agreement.

### SUMMARY

In conclusion, these results underscore the complexity and variability inherent in aspect-based sentiment analysis, particularly when applied to user comments which often exhibit high degrees of ambiguity and specific jargon. This inherent complication is reflected in the somewhat lower agreement rates among human coders, which, despite being fairly consistent, reveal the difficulties in achieving unanimous interpretations for the ABSA. The performance gap between humans and both LLM A and LLM B accentuates the challenge, as these models, while consistent, fail to reach the same level of agreement with the human coders. Furthermore, the differences between LLM A and LLM B emphasize how both the selection of the model and the chosen prompting strategies impact the results of ABSA. However, these differences do not allow for a clear winner in terms of prompting techniques, as the performances are too similar. At most, slight disadvantages can be identified for the CoT prompts.

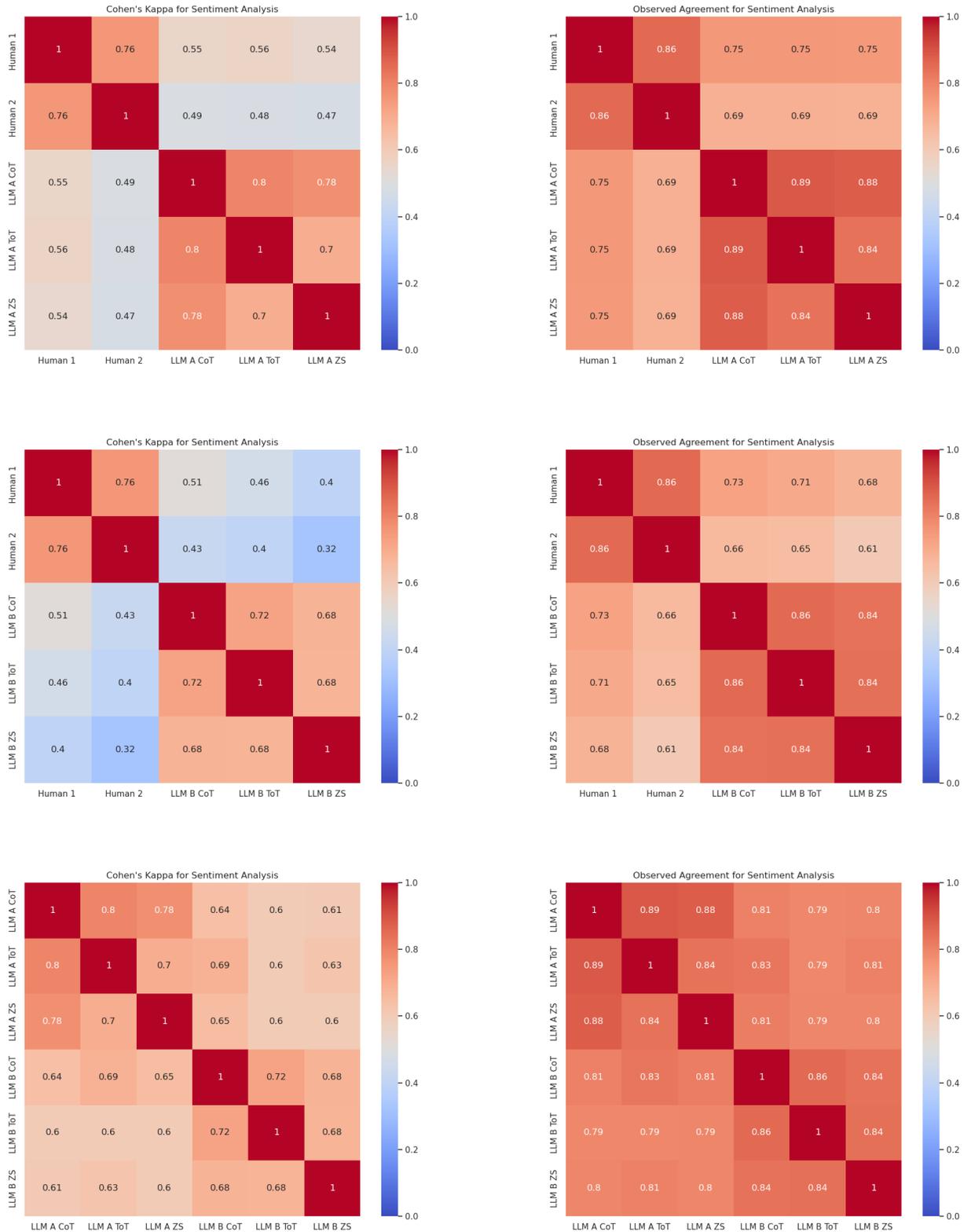

Figure 1: Cohen's Kappa and Observed Agreement between LLM Coding and Human Coding for ABSA

## RESULTS FOR GAMING EXPERIENCE

### HUMAN CODERS

The comparison among human coders, gauged using Cohen's Kappa, revealed a significant agreement score of 0.74. Supported by an observed agreement of 0.91. These results demonstrate consistent judgments by human coders, though they are not flawless, highlighting the inherent complexities in evaluating comments about the influence of lootboxes for the gaming experiences based on the user generated text data (see Figure 2). Consequently, it can be inferred that LLMs, like human coders, will likely encounter challenges in reaching elevated levels of agreement.

### LLM A

When comparing the agreement between human coders and LLM A, Cohen's Kappa values indicated moderate to fair agreement. Specifically, Human 1 and LLM A using CoT prompting showed a Kappa value of 0.55, while the agreements with ToT and ZS prompting methods were 0.51 and 0.36, respectively. Human 2 exhibited slightly lower Kappa values with LLM A: 0.48 for CoT, 0.43 for ToT, and 0.35 for ZS. These moderate Kappa values were supported by observed agreement values ranging from 0.79 to 0.85, suggesting that while the prompts of LLM A are fairly consistent with human coders, they do not achieve human-level agreement for the gaming experience analysis.

### LLM B

Similarly, LLM B showed moderate to fair agreement with human coders. Cohen's Kappa values were 0.36 for Human 1 vs. LLM B CoT, 0.33 for ToT, and 0.32 for ZS. Human 2 had slightly lower Kappa values: 0.29 for CoT, 0.33 for ToT, and 0.22 for ZS. These values were corroborated by observed agreement values ranging from 0.75 to 0.81. These results are clearly lower then LLM A, especially for the CoT and ToT prompts.

### LLM A VS. LLM B

The analysis of Cohen's Kappa values reveals substantial agreement within each LLM set and moderate agreement between LLM A and LLM B in the context of gaming experience analysis. Within LLM A, Kappa values range from 0.49 to 0.64, indicating high internal consistency across different prompting techniques. For LLM B, Kappa values range from 0.39 to 0.53, showing substantial consistency but slightly more variability compared to LLM A.

Comparing LLM A and LLM B, Kappa values range from 0.42 to 0.52, indicating moderate agreement. This suggests reasonable alignment between the two models, though with notable differences in their classifications. Observed agreement values support these findings, ranging from 0.81 to 0.88 within LLM A, 0.8 to 0.87 within LLM B, and 0.81 to 0.84 between the two LLMs.

In conclusion, while both LLMs demonstrate high internal consistency, there is moderate variability between them, highlighting the importance of model and prompting technique selection also in our second task of gaming experience analysis.

## SUMMARY

LLM A demonstrated moderate to fair agreement with human coders, with the CoT and ToT prompting showing the highest consistency. LLM B, although consistent, generally exhibited lower agreement levels compared to LLM A, particularly in CoT and ToT prompting. The ZS Prompt, on the other hand, show poor agreement values for both LLM A and LLM B.

Both LLMs showed internal consistency across different prompting techniques. However, it is important to note that all these agreements were lower than those observed between human coders. LLM A's Kappa values indicated substantial agreement within its prompts, while LLM B's values reflected slightly more variability. The moderate agreement between LLM A and LLM B highlights reasonable alignment but also notable differences in classifications, emphasizing the impact of model selection. Consequently, both the internal and cross-LLM agreements are relatively low compared to the agreement between human coders.

Overall, it can be stated that the assessment of whether the text expresses an influence of lootboxes etc. on the gaming experience is a similar or even more difficult task for the LLMs than the ABSA.

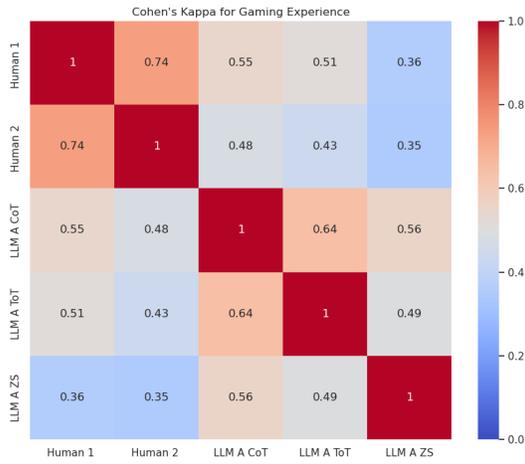
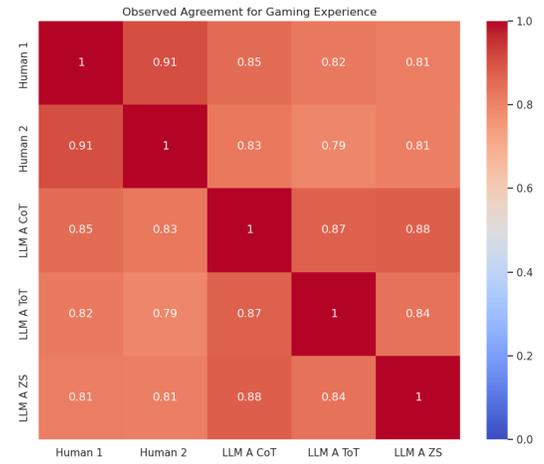
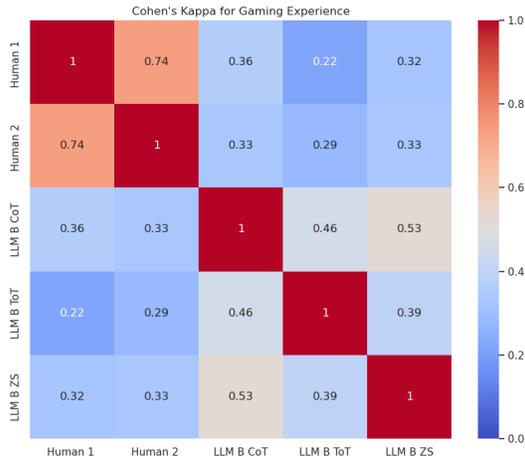
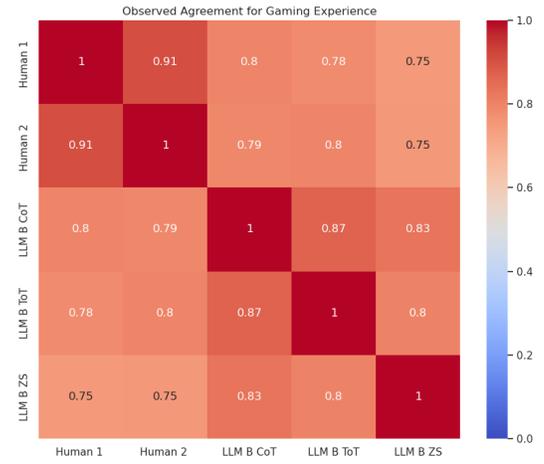
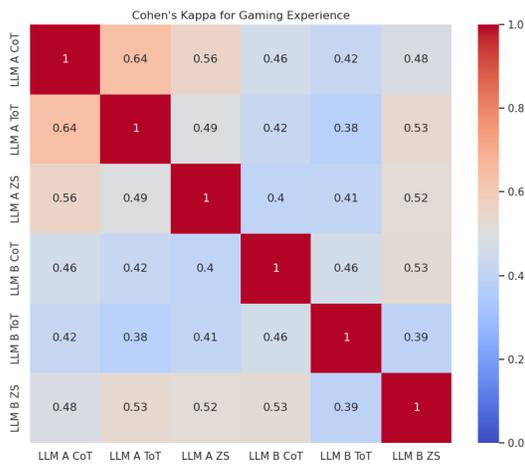
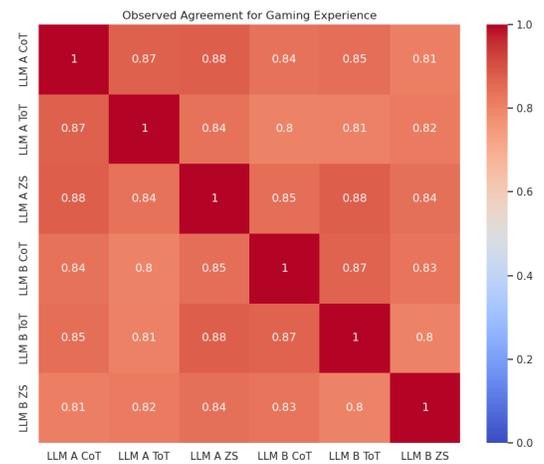

**Figure 2: Cohen's Kappa and Observed Agreement between LLM Coding and Human Coding for Gaming Experience**

## RESULTS FOR FINANCIAL ENGAGEMENT

### HUMAN CODERS

The comparison among human coders for payment evaluation showed high consistency, with Cohen's Kappa values of 0.91 between Human 1 and Human 2. This is further supported by an observed agreement of 0.98, indicating that human coders have a very high level of agreement in their classifications. This strong agreement among human coders reflects their ability to consistently evaluate comments about payment aspects in user-generated text data.

### LLM A

When assessing the agreement between human coders and LLM A, Cohen's Kappa values indicated substantial agreement. For instance, the Kappa value between Human 1 and LLM A CoT was 0.73, while for ToT and ZS prompting methods, the values were 0.69 and 0.77, respectively. Similarly, Human 2 showed Kappa values of 0.7 for CoT, 0.66 for ToT, and 0.73 for ZS. These substantial Kappa values were supported by observed agreement values ranging from 0.91 to 0.94, suggesting that LLM A's prompts align closely with human coders' classifications.

### LLM B

LLM B showed fair to substantial agreement with human coders. The Kappa values for Human 1 were 0.71 for CoT, 0.68 for ToT, and 0.6 for ZS, while for Human 2, the values were 0.67 for CoT, 0.65 for ToT, and 0.58 for ZS. These Kappa values indicate that while LLM B is consistent, its agreement with human coders is somewhat lower than that of LLM A, especially for the ZS prompt. The observed agreement values ranged from 0.86 to 0.93, reflecting similar trends.

### LLM A VS LLM B

The internal consistency within each LLM set was notable, particularly for LLM A, with Kappa values between prompting techniques ranging from 0.79 to 0.88. LLM B showed Kappa values from 0.57 to 0.81, indicating more variability within its prompts. The observed agreement within LLM A ranged from 0.94 to 0.97, while within LLM B, it ranged from 0.88 to 0.95.

When comparing LLM A to LLM B, the Kappa values ranged from 0.55 to 0.81, indicating moderate to high agreement. The observed agreement values between LLM A and LLM B ranged from 0.84 to 0.95, showing that while they generally align, the agreement is not as high as within each LLM set. In particular, the ZS variant of LLM B generates the wide range of approval values.

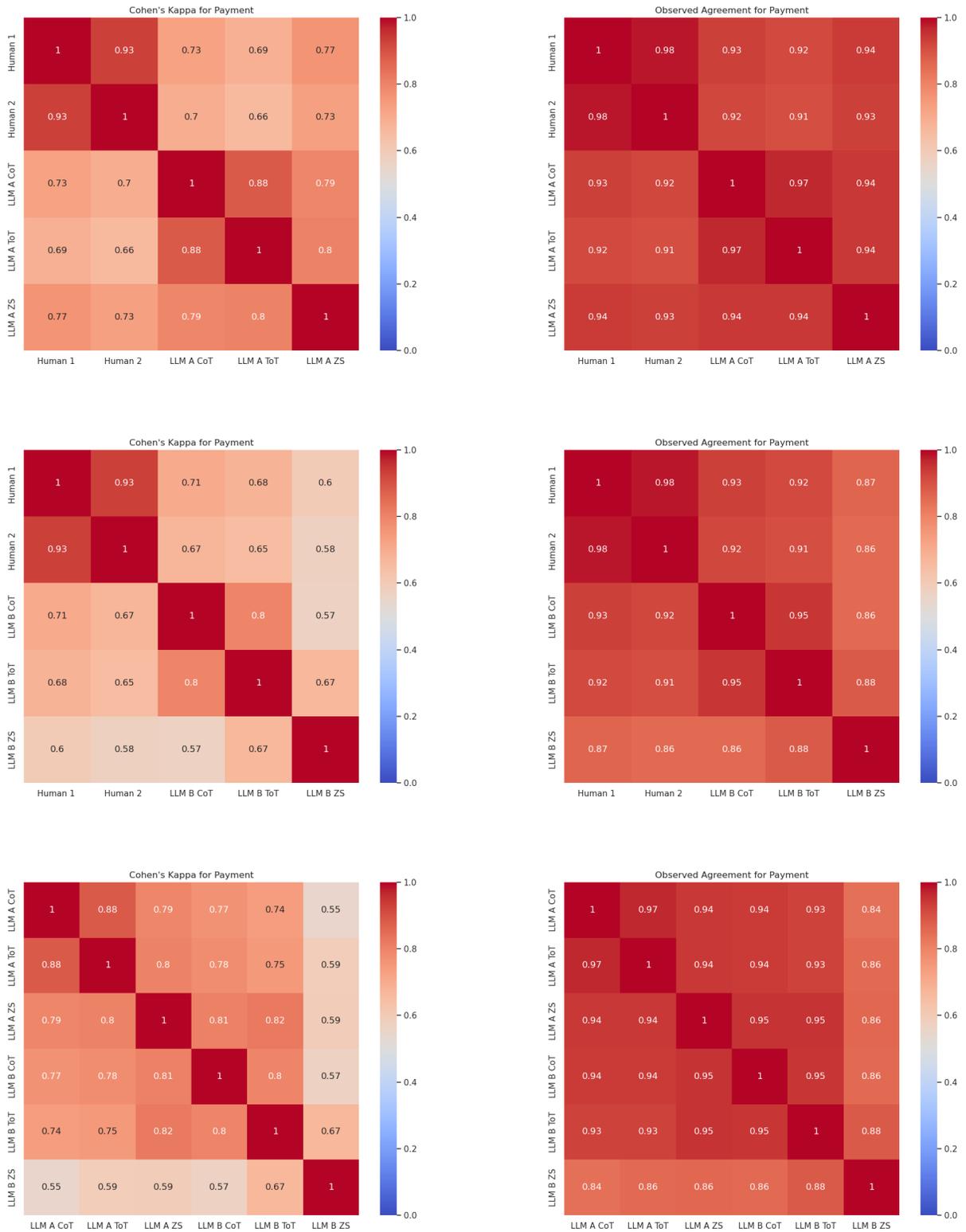

Figure 3: Cohen's Kappa and Observed Agreement between LLM Coding and Human Coding for Financial Engagement

## SUMMARY

As the high level of agreement between the human coders has already shown, classifying whether a user text explicitly states whether money was spent on lootboxes etc. is a comparatively simple classification task.

LLM A demonstrates substantial agreement with human coders and high internal consistency, while LLM B, although consistent, shows more variability and lower agreement with human coders compared to LLM A, which is mainly driven by the bad performance of the ZS variant. The partly moderate agreement between LLM A and LLM B highlights the importance of model selection and prompting techniques, as they can have a significantly impact the consistency and reliability of classification. In terms of prompting techniques, however, there is no clear winner. Although ZS shows less agreement with LLM B, it actually proves to be the best variant with LLM A.

Overall, it can be stated that for the classification of the financial engagement dimension, the classification performance of the LLMs almost reaches the level of human coders.

## RESULTS FOR GAMBLING COMPARISON

### HUMAN CODERS

The agreement between human coders is exceptionally high, with Cohen's Kappa at 0.93 and observed agreement at 0.98. This indicates a strong consistency in human evaluations of gambling experiences. This near-perfect agreement among human coders suggests that the classification task is relatively straightforward and clear, implying that LLMs should also perform well in this task.

### LLM A

When comparing the agreement between human coders and LLM A, Cohen's Kappa values indicated near-perfect agreement. For instance, the Kappa value between Human 1 and LLM A using CoT prompting was 0.9, while agreements with ToT and ZS prompting methods were 0.82 and 0.86, respectively. Human 2 exhibited similar agreement levels, with Kappa values of 0.89 for CoT, 0.87 for ToT, and 0.82 for ZS. These values are supported by observed agreement values ranging from 0.95 to 0.97, demonstrating that LLM A's prompts closely align with human coders' evaluations. According to Landis and Koch (1977), these values can be classified as "almost perfect agreement" for all prompting techniques.

### LLM B

Similarly, LLM B showed substantial to near perfect agreement with human coders, though slightly lower than LLM A for the ToT and CoT prompts. The Kappa values range from 0.72 to 0.92, with the ZS prompt showing the highest correspondence with the human coders.

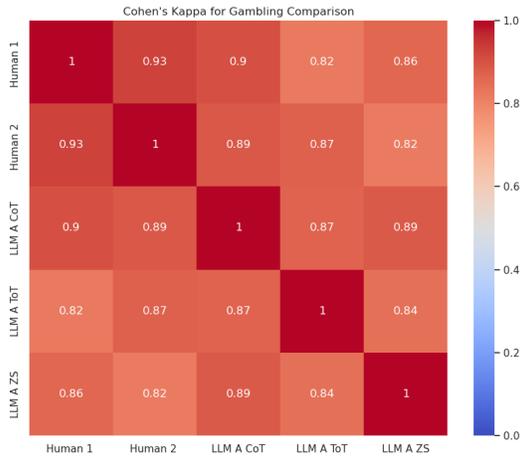
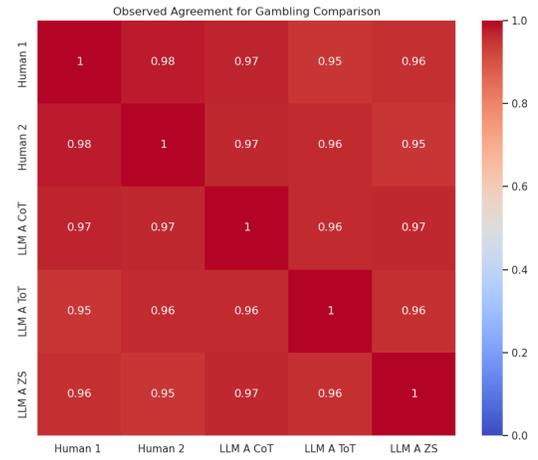
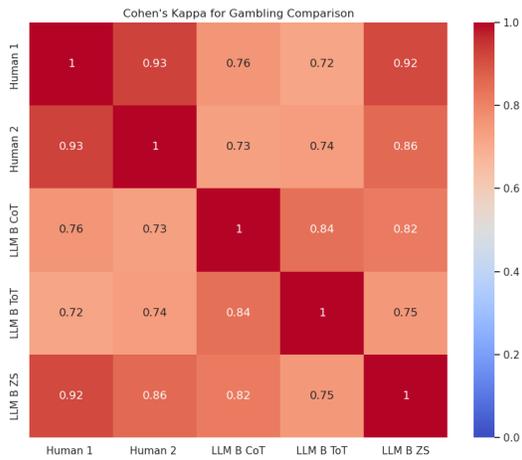
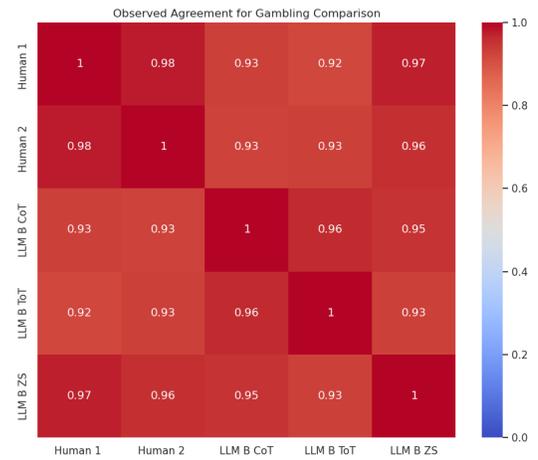
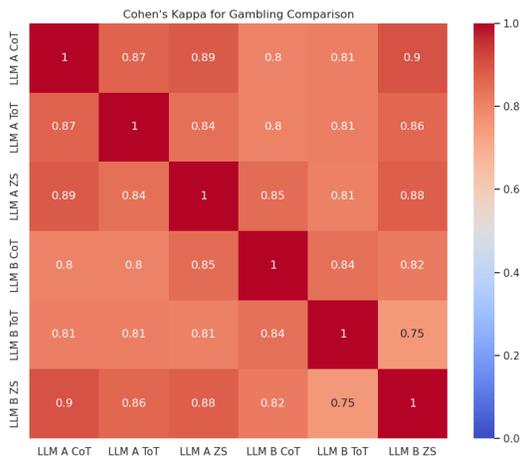
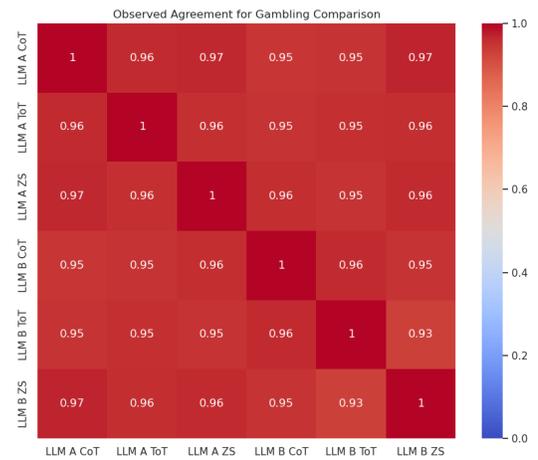

**Figure 4: Cohen's Kappa and Observed Agreement between LLM Coding and Human Coding for Gambling Comparison**

### LLM A VS LLM B

When comparing internal consistency within each LLM set, LLM A showed Kappa values between prompting techniques ranging from 0.84 to 0.89, indicating high internal consistency. LLM B demonstrated Kappa values ranging from 0.75 to 0.84, showing substantial but more variable internal consistency. Notably, the ZS prompt of LLM B is much more closely aligned with the results of LLM A's prompts than with LLM B's other variants.

### SUMMARY

Overall, it can be stated that the classification of whether the user texts compare lootboxes and similar mechanisms with gambling is a comparatively easy task for both the human coders and the LLMs. LLM A and LLM B (with ZS) achieve a degree of correspondence with the human codes that can be classified as equivalent. This high level of agreement underscores the effectiveness of both models in handling this classification task.

## RESULTS FOR ALL USER TEXTS

After demonstrating that the classification for Financial Engagement and Gambling Comparison nearly reaches the level of human coders, we applied the Large Language Models to our entire dataset of user comments. Initially, the LLMs filtered out comments that mentioned lootboxes and similar topics.[4] Subsequently, the best-performing prompts for the Financial Engagement and Gambling Comparison dimensions were applied to these filtered comments.

Our analysis included a total of 138,439 user comments. Among these, 1,117 specifically mentioned lootboxes and similar mechanics, accounting for only 0.81% of the total. This indicates that while lootboxes are a topic of discussion, they are not predominant in user comments. This low percentage is consistent with the expectation for player reviews on platforms like Steam. Most players tend to focus their reviews on the overall gaming experience, including gameplay mechanics, graphics, story, and performance issues. Lootboxes, although a significant concern for some players, often do not dominate the discourse. Therefore, the relatively low proportion of comments mentioning lootboxes aligns with the broader patterns observed in gaming communities, where general gameplay feedback is usually the primary focus.

These 1,117 user texts, where analysed by our best performing best-performing prompts.

According to LLM A (ZS), 84.24% of comments did not discuss Financial Engagement, whereas 15.76% did. Similarly, LLM B (CoT) showed that 85.04% of comments did not mention spending money on lootboxes, while 14.96% did.

The analysis for LLM A (CoT) showed that 89.34% of the user text did not make comparison between lootboxes and gambling, while 10.66% did. This was almost identical for LLM B (ZS), with 89.33% of comments not mentioning gambling and 10.67% doing so.

---

[4] To achieve this, a single-shot prompt was utilised with ChatGPT4-turbo and GPT4o-mini, achieving a classification agreement rate of 99.7%. In instances of conflicting classifications, the cases were designated as lootbox-relevant text.

These results demonstrate a high level of consistency between the two models in identifying gambling comparisons and identifying comments that mention spending money on lootboxes comments (see also Figure 5). The overall findings indicate that comments about lootboxes, while present, constitute a small fraction of the overall discussion. The models used show high consistency in their classifications, which is crucial for reliable text classification.

It is important to note that these percentages should not be misinterpreted as a measure of user opinions or a survey of purchasing behaviour. Rather, they represent the distribution of themes discussed in user comments. This analysis offers a window into the digital discourse, capturing what topics are being debated and the prevalence of these discussions within the broader context of user feedback on their gaming experiences. By examining the thematic distribution, we gain valuable insights into the focal points of user discussions, which reflect the digital conversations rather than direct measures of user attitudes or behaviours.

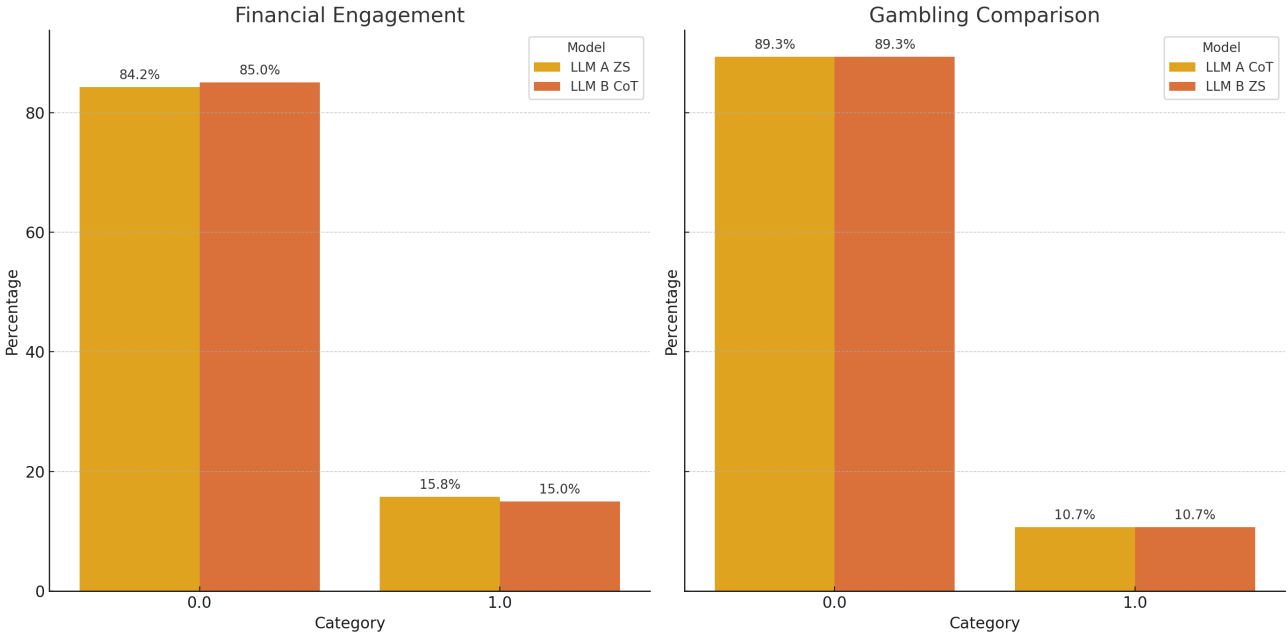

Figure 5: Percentage of mentions of Financial Engagement and Gambling Comparison in user comments addressing lootboxes etc. N = 1117.

### RELIABILITY OF RESULTS FOR ALL USER TEXTS

After implementing the four best performing models and prompts across the dataset for a descriptive picture of the used themes, our next objective was to assess the robustness and reliability of the LLM classifications. While the validity of these models was previously assessed through comparison with human coders, the focus now shifts to examining their reliability.

To achieve this, we deployed 10 separate LLM instances for each of the four prompt-model combinations, providing identical input from our dataset of 1,117 user-texts mentioning lootboxes under controlled conditions. We set the temperature parameter for each instance to 0.0, aiming for quasi-deterministic outputs that should result in near-perfect reliability. To evaluate the reliability of our LLM classifications, we will employ Krippendorff's Alpha as a statistical measure of agreement among the classifications produced by the 10 deployed LLM instances. This coefficient is used for our study due to its capacity to handle data from multiple coders, making it ideal for assessing consistency across the instances under controlled conditions. Krippendorff's Alpha also corrects for chance agreement, offering a more accurate depiction of reliability than simpler metrics. This robust method will allow us to ascertain whether the quasi-deterministic outputs of the LLMs, achieved by setting the temperature parameter to 0.0, hold true.

The results were as follows (also compare Figure 6):

Alpha for LLM A Zero-Shot in Financial Engagement was 0.9297, with a 95% confidence interval of (0.9090, 0.9474).[5] This indicates a strong level of agreement among the LLM instances for this prompt-model combination, though there were 55 instances out of the 1,117 user texts where some disagreement occurred among the LLM coder instances.

Alpha for LLM A Chain-of-Thought in Gambling Comparison was higher at 0.9599, with a 95% confidence interval of (0.9408, 0.9761). This suggests even greater reliability in the classifications, with only 23 instances out of 1117 user texts showing disagreement among the LLM instances.

Both LLM B Zero-Shot in Gambling Comparison and LLM B Chain-of-Thought in Financial Engagement achieved an Alpha of 1.0, indicating perfect agreement across all LLM instances.

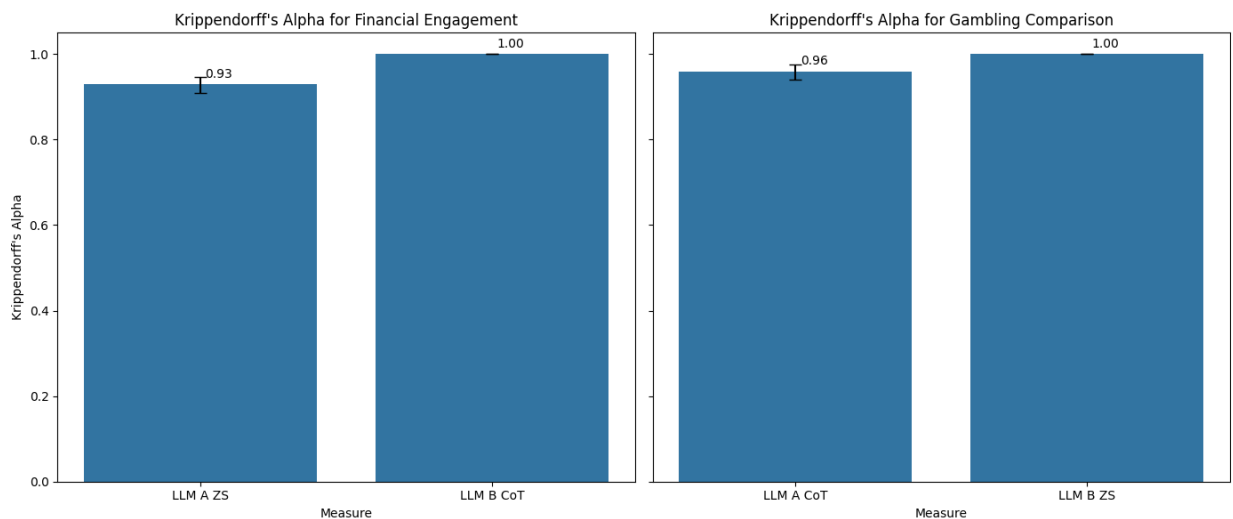

Figure 6: Reliability of LLM Classifications of 10 LLM Instances of 1,177 user texts

These results demonstrate that while LLM A exhibits some degree of variability in its responses, even with a temperature setting of 0.0, this randomness is minimal and does not pose significant practical issues for our application. The low levels of disagreement suggest that the quasi-deterministic nature of the outputs is sufficient for reliable classification. Notably, LLM B demonstrated more uniform codings at a temperature setting of 0.0, showing no variability in its classifications, which led to consistent and deterministic classification outputs across all cases.

---

[5] The confidence intervals for Krippendorff's Alpha were each calculated using 1,000 bootstrapped samples.

Overall, these findings confirm that under the controlled conditions of this study, the LLM models produce highly reliable classifications. The high Krippendorff's Alpha values highlight the effectiveness of using a temperature setting of 0.0 to achieve near-deterministic and reproducible outputs across multiple instances. Consequently, these models can be confidently applied to provide robust insights.

# DISCUSSION

## TASK PERFORMANCE

The performance of large language models across the four tasks investigated in this study—aspect-based sentiment analysis, gaming experience, financial engagement, and gambling comparisons—revealed significant differences in model capabilities. The models demonstrated varied success, with their performance being generally lower compared to human coders in more complex analytical tasks such as aspect-based sentiment analysis and gaming experience. However, in more straightforward tasks such as financial engagement and gambling comparisons, the models approached or matched human-level performance.

**Aspect-Based Sentiment Analysis (ABSA):** In the task of ABSA, both LLMs, though consistent, did not achieve the level of agreement with human coders observed among the human coders themselves. The moderate agreement levels, as indicated by Cohen's Kappa, suggest that while the LLMs are capable of understanding and classifying sentiments, the finer nuances and ambiguity of language and context present in user texts remain challenging. This is at least true if human ratings are taken as the basis for evaluating performance. The different prompting methods—Chain-of-Thought (CoT), Three-of-Thought (ToT), and Zero-Shot (ZS)—showed that more structured prompting (CoT and ToT) generally leads to slightly better performance than ZS, underscoring the importance of tailored prompting in enhancing model accuracy.

**Gaming Experience:** Evaluating the influence of lootboxes on gaming experience proved challenging for the models. The task's inherent subjectivity and the nuanced nature of gaming terminology may contribute to lower agreement rates with human coders. Here again, the CoT prompting method provided marginally better results than ToT and ZS, suggesting that guiding the models to reason through the task step-by-step could improve their output quality.

**Financial Engagement:** Both LLMs showed substantial agreement with human coders in identifying financial transactions related to in-game purchases of lootboxes. This task, being more straightforward and often involving specific keywords, aligns well with the capabilities of LLMs, allowing them to achieve near-human performance. This suggests that LLMs are well-suited to tasks with clear, definable criteria and less linguistic ambiguity.

**Gambling Comparisons:** The performance of large language models in the task of identifying comparisons between lootbox mechanics and traditional gambling activities showcased some of the most striking capabilities of these models. Notably, both LLMs are able to exhibit substantial to near-perfect agreement scores with human coders. This high level of performance offers important insights into the strengths of LLMs when dealing with explicit and straightforward content without the need to decide on linguistic ambiguity.

Although the goal was to achieve human-like performance, the LLMs demonstrated relatively high accuracy ratings, which were evident from the observed agreements with human coders. These substantial observed agreement values, often with only about a 10% accuracy deviation from human coders even for the complex tasks, indicate a robust level of performance. Even though these models did not completely match the level of agreement seen among human coders, their performance is notable, especially given the inherent ambiguities and complexities in user-generated texts. In contexts involving large datasets, such as extensive textual data or social media texts, this level of accuracy may already be acceptable. While LLMs may not achieve perfect accuracy for every task, their ability to handle large datasets efficiently, consistently, and cost-effectively makes them valuable tools in modern data analysis. The acceptance of suboptimal performance is balanced by the significant benefits they offer in terms of scalability, resource optimization, and the ability to derive meaningful insights from vast amounts of data. Our additional analyses of classification reliability at a temperature setting of 0.0 further demonstrate that the LLM models consistency. They are not only accurate but also consistently stable in their classifications. The high Krippendorff's Alpha values illustrate the effectiveness of this setting in achieving outputs that are nearly deterministic and reproducible across multiple instances. Consequently, these models can be applied with confidence to deliver reliable and insightful results.

## PROMPTING TECHNIQUES

When considering the different prompting techniques—Chain-of-Thought (CoT), Collaborative Reasoning/Three-of-Thought (ToT), and Zero-Shot (ZS)—used across the tasks, it becomes apparent that there are discernible differences in response quality. Each technique, tailored to exploit specific characteristics of large language models, has its strengths and weaknesses. However, across all target dimensions and models, no single prompting technique consistently outperforms the others.

For instance, in the aspect-based sentiment analysis and gaming experience tasks, CoT generally delivered slightly superior outcomes, likely due to its ability to guide models through a more structured thought process, thereby enhancing the clarity and relevance of the responses. his structured approach proves advantageous in tasks that require a deep understanding of context and the ability to navigate linguistic subtleties. On the other hand, the ToT method, which encourages a collaborative reasoning process, did not consistently yield better results compared to CoT and sometimes even performed similarly to ZS, depending on the specific task and context.

Zero-Shot prompting, while often less effective in highly nuanced tasks, showed comparable or even superior performance in straightforward tasks such as financial engagement and gambling comparisons. In these cases, the inherent capabilities of the models to draw on their extensive pre-trained knowledge without specific guidance proved sufficient and effective. This highlights that the effectiveness of a prompting technique can largely depend on the nature of the task—more structured prompting seems beneficial for complex, nuanced tasks, while simpler, direct tasks may not require such detailed guidance.

This variance underscores the complexity of working with LLMs and suggests that the choice of prompting technique should be carefully considered based on the specific requirements of the task at hand. While the results indicate that no one method is universally superior, they do highlight the importance of matching the prompting strategy to the task's characteristics to optimize performance. In examining the effectiveness of the Chain-of-Thought and Three-of-Thought prompting techniques in the context of text classification, particularly with ambiguous text, it becomes evident that these methods do not provide the same advantages as they do in logic-based tasks with a definitive right answer. While CoT and ToT are designed to enhance model reasoning by guiding it through a sequence of logical steps, their application in text classification tasks involving ambiguous language and nuanced or unclear sentiment does not consistently yield the expected improvements. This discrepancy can be attributed to the inherent challenges posed by the subjective interpretation of ambiguous text, where multiple valid interpretations can exist, making it difficult for models to follow a linear, logical path to a definitive conclusion, which will also align with the human coders.

In tasks that require the disambiguation of complex language or the detection of subtle emotional nuances, the structured nature of CoT and ToT may not align perfectly with the demands of the task. These methods, although beneficial in breaking down complex reasoning into simpler, sequential steps, assume a level of clarity and objectivity that ambiguous text often lacks.

During the iterative prompt development process, we observed that the arguments provided by the models were generally acceptable, comprehensible and generally logical, even when they did not coincide with human evaluations. No amount of additional argumentative depth (CoT) or additional expert opinions (ToT) can bridge this gap. Consequently, while CoT and ToT can guide models through structured reasoning in logic-based scenarios, their utility is somewhat diminished in the face of textual ambiguity where flexibility and a deeper contextual understanding are required. This highlights the need for further refinement of prompting techniques to better handle the complexities of ambiguous language, potentially incorporating more dynamic and context-aware strategies that can adapt to the inherent variability of natural language.

### REAL-WORLD TEXT CHALLENGES

To address the nature of the user text data involved in our study, it is crucial to recognize the broad spectrum of formality, slang, and motives that these texts encompass, presenting a particularly challenging data situation. User-generated content, by its very nature, varies widely in style, context, and clarity, often diverging significantly from the structured and well-defined datasets typically used in training and tuning large language models. This variability includes casual expressions, slang, and highly personalized expressions of opinion, all of which contribute to a complex linguistic landscape that LLMs must navigate.

One illustrative but still simple example from our dataset is the user comment:

*"sold my soul for cases. would recommend."*

This comment epitomizes the challenges inherent in sentiment analysis due to its ambivalent nature. The interpretation of this comment can justify multiple sentiment classifications. The phrase "sold my soul" typically conveys a sense of significant personal sacrifice or regret, suggesting a negative experience. Conversely, the phrase "would recommend" implies satisfaction and approval, indicating that despite the sacrifice, the user finds the overall experience worthwhile and endorses it. The juxtaposition of a regretful sacrifice with a positive recommendation captures both positive and negative emotions, reflecting the duality of the user's experience, which could also justify a "mixed" classification. The crucial point is that for each categorization, strong arguments can be (and are) made, all of which are fundamentally reasonable, yet it ultimately remains indeterminable which classification is correct.

Another example from the text data exemplifies the difficulties for the gaming experience classification:

*"[...] There's always some RNG when it comes to shooting, but cs2 is the only game I've ever played where shooting a gun or going around a corner feels like opening loot crates. Being rewarded seemingly at random is not as fun as having some consistency even if being consistent means being the bottom of the score board. [...]"*

The model must recognize that the user is drawing a parallel between the randomness in core gameplay mechanics (shooting and movement) and the randomness associated with opening loot crates. This requires the model to interpret "feels like opening loot crates" as a critique of the unpredictability in these gameplay elements. However, the user is making a comparison rather than stating that lootboxes directly influence these gameplay elements. This distinction is crucial yet subtle, and it presents a significant challenge for large language models). The model needs to understand that the user's comparison highlights a perceived similarity in the experience of randomness, without implying that the mechanics of lootboxes have directly altered the shooting or movement mechanics in the game. The comment suggests that the randomness of shooting and movement feels like the randomness of lootboxes, but it does not claim that lootboxes are influencing these core gameplay elements. LLMs can struggle with such nuanced distinctions. They might overinterpret the comparison, mistakenly categorizing the comment as indicating that lootboxes have a direct influence on gameplay mechanics.[6]

Given these complexities, the performance of the LLMs under study might have been constrained by the inherent challenges posed by such diverse and informal text data. Even the human coders were faced with interpretative challenges, as reflected in the deviations in their assessments. It is conceivable that under different conditions—such as with texts that exhibit more uniformity in terms of language use and thematic content—the results for the LLMs could have been markedly better. This consideration underscores the potential variability in LLM performance and highlights the importance of context and data characteristics in evaluating these models. Therefore, it is reasonable to suggest that the outcomes observed in this study are not definitive limits of the models' capabilities but rather reflections of the particularly challenging nature of the data used.

Moreover, the challenging nature of the data used in our study can also be viewed as a strength of our research. By utilizing texts directly from the "wild" online environment—without relying on standardized or pre-classified texts—we embraced the inherent difficulties and imperfections that such data entail. This approach has enabled us to assess the performance of large language models under real-world conditions, providing a measure of external validity that closely mirrors the actual complexities and nuances encountered in everyday online interactions. This methodological choice is significant because it moves beyond the often-idealized conditions typical of many computational linguistics studies, which tend to use cleaner, more controlled datasets. By tackling a diverse and realistic dataset, our study offers insights into how LLMs perform when they are applied in settings that they are likely to encounter outside of standardized conditions. This is crucial for understanding the practical implications of deploying these models in real-world applications, where the data is rarely clean or predictable. Thus, our approach not only tests the robustness of LLMs but also provides a more accurate reflection of their utility and limitations in everyday use, contributing significantly to the ongoing dialogue on the practical deployment of artificial intelligence in natural language processing.

---

[6] We could extend the discussion by enumerating many more additional edge cases and ambiguities that contribute to the misalignments between human interpretations and LLM outputs. However, for the sake of brevity, we limit our discussion to these two examples.

## COMPARISON WITH EXISTING LITERATURE

Our study extends the findings of Zhang et al. (2023), who conducted an investigation into the capabilities of LLMs in sentiment analysis. Zhang et al. evaluated LLMs across a range of sentiment analysis tasks, including both traditional sentiment classification and more complex challenges such as aspect-based sentiment analysis. They faced limitations with more intricate tasks requiring deeper text understanding. In our study, we observed similar patterns. For straightforward tasks such as identifying financial engagement and recognizing gambling comparisons, the LLMs achieved near-human performance levels. However, in more complex tasks like aspect-based sentiment analysis and evaluating effects on the gaming experience, the models' performance was less impressive. Our findings reinforce the conclusions of Zhang et al., emphasizing the need for continued refinement of LLM capabilities to fully realize their potential in sentiment analysis. In the overall comparison, we can also confirm that a higher number of categories generally makes the classification task more difficult, as can be inferred from the performance for ABSA studies (Wang and Luo 2023; Krugmann and Hartmann 2024).

Regarding the mixed results concerning the performance our prompting strategies, our findings parallel those of Kuila and Sarkar (2024), who also found no consistent higher effectiveness for CoT Prompting in their analysis of sentiment for news entities. Our results regarding prompting strategies somewhat contradict the interpretations of Wang and Luo (2023), who were able to achieve better results with CoT and a combination of roleplaying and CoT. However, it must be noted that this better performance in Wang and Luo (2023) was only valid to specific subsets of their data and was conducted using an earlier version of GPT (3.5). This substantiates the interpretation that performance depends on the underlying Text-Dataset and is not universally valid. Additionally, our study aligns with Reiss (2023), who highlighted that even minor wording alterations in prompts can lead to varying outputs, which speaks to the need for iterative prompt development and underlines the variable performance of different prompting strategies.

To a certain extent, our findings also are in line with the results of Krugmann and Hartmann (2024), who were able to identify poor classification performance of LLMs for short and unstructured social media posts, which they identified using a standard sentiment analysis.

## EXCURSION: CHALLENGES IN MODEL RESPONSES AND ETHICAL CONSTRAINTS

In the course of our study, we encountered specific types of user-generated tests that impacted the response quality of our models and the effectiveness of the prompting techniques used. Notably, texts containing racial slurs, repetitive characters, and mentions of suicidal thoughts posed distinct challenges.

When processing texts with racial slurs, the models (mainly LLM B) often deviated from the provided instructions, by producing responses that lacked relevance to the task at hand.[7] This deviation likely stems from the built-in safety and ethical guidelines embedded in the models, designed to prevent the dissemination of harmful or offensive material. While these safeguards may be crucial for responsible AI deployment, they can interfere with the models' ability to follow specific analytical instructions in a controlled research environment.

Texts composed entirely of repetitive characters (e.g., "aaaaaaa" or "!!!!!!!") frequently caused the models to produce errors or entirely ignore the provided instructions. For example, when faced with the repetitive sequence of "skyskysky," one model (LLM A) generated an output text about the U.S. presidential election in 2014, completely disregarding the original task. Such inputs sometimes also led to exceptions being thrown during API calls, disrupting the analysis process. These types of texts present a form of noise that the models are not well-equipped to handle, leading to breakdowns in the standard processing flow and highlighting a limitation in the robustness of current LLMs when faced with non-standard inputs.

Moreover, texts that included mentions of suicidal thoughts or offensive language triggered significant deviations in the models' responses. The models (mainly LLM B) sometimes prioritized addressing the sensitive nature of the content over following the original classification instructions, sometimes producing generic crisis intervention messages or refused to help with a task that includes offensive language. This behaviour reflects the ethical programming intended to mitigate potential harm but also illustrates a critical challenge in maintaining analytical accuracy in the presence of sensitive topics.

These issues underscore the importance of refinement during prompt engineering and basic quality control of the data to better accommodate and filter out such problematic inputs or react to unexpected responses. Doing so enhance the reliability and consistency of LLM outputs, particularly in diverse and uncontrolled environments like online user-generated content. While LLMs generally do not require extensive data pre-processing and can often even recognize ASCII art, or emotes they sometimes struggle to handle such unconventional inputs appropriately.

Furthermore, we found that the adherence to the more elaborate prompt variants varied between the two models. The CoT and ToT prompts did not always yield consistent results across both models. Consequently, minor prompt adjustments[8], such as reminders of the required procedure, were necessary to guide the models consistently. These supplementary instructions helped improve adherence to task-specific guidelines, highlighting the need for flexibility and adaptability in prompt engineering especially if different LLMs are used. Each model can exhibit its own unique characteristics and does not always respond identically to the same prompts. While the outputs or task adherence may be consistent for most the inputs, various degrees of deviation can occur in a small number of cases. This variability underscores the inherent unpredictability of these "black box" models, emphasizing the importance of ongoing evaluation and adjustment to ensure reliable performance across different tasks and contexts.

---

[7] Krugmann and Hartmann (2024) also observed this behavior for the predecessor model Llama2 in the context of sentiment analyses.

[8] E.g. „Remember to think step by step." for CoT or "Proceed to provide the complete analysis without stopping. You must give a final answer code at the end of your answer." for ToT.

## IMPLICATIONS AND RECOMMENDATIONS

Large language models such as ChatGPT and Llama have raised hopes of making advanced text analysis accessible to a broader audience of researchers without requiring specialized knowledge, extensive pre-processing of text data, or model training and fine-tuning. These models promise to democratize text analytics, enabling individuals across various disciplines to harness the potential of LLMs for their research and applications.

While LLMs show promise in understanding context, including language with humour and sarcasm, and community specific slang, they still struggle with ambiguity. Interestingly, during the iterative prompt development, we could observe that LLMs occasionally recognized domain-specific jargon better than human coders, highlighting their potential in specialized contexts. Despite advancements, the inherent complexity and subtlety of human language mean that fully resolving ambiguities still remains a challenge. This limitation underscores the need for continued development and refinement of the use of these models to enhance their alignment with human understanding capabilities. It is important to note that ambiguity in text is also a challenge for human coders. Therefore, achieving human-level performance for LLMs means achieving a high degree of alignment with human interpretations. This involves understanding and mimicking the nuanced decision-making processes of human coders and trying to explicitly formulate them in form of a prompt for the LLMs.

So, iterative prompt development remains crucial for achieving the best possible results with LLMs. Crafting and refining prompts through an iterative process allows for better alignment with the task requirements and can significantly improve the model's performance. This iterative approach is essential to tailor the model's responses to specific analytical needs. However, it is not always necessary to start from scratch. Studies like this one provide a foundation for prompt development that can be further refined and adapted to fit specific research interests.

Achieving text classification at near-human levels appears feasible for certain types of tasks. However, this level of performance still requires substantial preliminary work in prompt development and quality assurance. There is no "magic button" solution that will work for all kinds of data and topics; effective use of LLMs involves careful design, testing, and refinement to ensure accuracy and reliability. This highlights the ongoing necessity for human expertise and intervention to optimize the capabilities of LLMs in text classification tasks. Therefore, comparing these results with human-generated classifications remains essential, as the unsupervised use of LLMs is generally not (yet) recommended (Reiss 2023).

Our analysis has shown that different prompting techniques can significantly influence responses and the performance of the responses generated by LLMs. In our study, we explored techniques such as Chain-of-Thought (CoT), Three-of-Thought (ToT), and Zero-Shot (ZS) prompting. These methods were particularly employed to test and enhance the reasoning capabilities of the models. By guiding the models through structured thought processes (CoT and ToT), we aimed to evaluate and improve their ability to perform complex reasoning tasks and better align their outputs with human interpretations. However, while these approaches may be effective for testing reasoning skills, they may not be the optimal strategy for text analysis tasks. Other approaches, such as persona prompting, where the model is guided to respond from the perspective of a specific persona or role, might be necessary. These alternative prompting methods could provide a framework to improve response relevance and alignment with human coders, further optimizing the models' performance across various tasks.

It is important to note that LLMs are continuously being developed and improved. Within a span of just a few months, newer and more powerful variants become available, offering enhanced capabilities and performance. This rapid advancement means that the tools and methods for text analysis are constantly evolving, and researchers can expect to access increasingly sophisticated and affordable models that can handle more complex or nuanced tasks and maybe better mimic human decision-making processes. Keeping abreast of these developments is crucial for leveraging the full potential of LLMs in research and practical applications.

## CONCLUSIONS AND FUTURE WORK

Our study has provided significant insights into the application of large language models for analysing user-generated text data, particularly in the context of gambling-like elements in digital games. The key findings from our research can be summarized as follows:

- **Performance Variability Across Tasks**: The LLMs demonstrated varying degrees of success across different tasks. They performed relatively well in straightforward tasks such as identifying financial engagement and recognizing gambling comparisons with lootboxes and similar mechanics, achieving near-human levels of agreement. However, their performance was less impressive in more complex analytical tasks such as aspect-based sentiment analysis (ABSA) and evaluating whether the gaming experience was influenced.
- **Influence of Prompting Techniques**: Our investigation into different prompting techniques—Chain-of-Thought (CoT), Collaborative Reasoning (ToT), and Zero-Shot (ZS)—revealed that no single technique consistently outperformed the others across all tasks. While CoT generally provided better results for complex tasks, ZS proved effective for simpler tasks, emphasizing the need for context- and task-specific prompt engineering.
- **Real-World Text Challenges**: The diverse and informal nature of user-generated content posed significant challenges for the LLMs. Variability in style, context, and clarity, including casual expressions and slang, complicated the models' ability to consistently categorize and interpret the data. Despite these challenges, our study underscores the importance of evaluating LLM performance in real-world conditions to understand their practical utility and limitations.
- **Model Specify**: Our study highlighted the need for flexibility and adaptability in model selection and prompt engineering. Each model exhibited unique characteristics and did not always respond identically to the same prompts, suggesting that prompt adjustments and exception handling are necessary to guide specific models effectively across different tasks.

To support future research, we are making our prompts and analysis scripts available to other researchers who are working on similar topics. By sharing these resources, we aim to provide a foundation that others can build upon, refine, and adapt to their specific research interests and needs. This collaborative approach will help advance the field of text analysis using LLMs and contribute to the development of more robust and effective analytical methodologies.

## FUTURE WORK

Our study has laid a foundation for the application of large language models in analysing user-generated text data, particularly in the context of gambling-like elements in digital games. However, there remain several avenues for future research and development that could further enhance the efficacy and applicability of these models.

While our investigation into various prompting techniques, such as Chain-of-Thought (CoT), Three-of-Thought (ToT), and Zero-Shot (ZS) prompting, has provided initial insights, there remains considerable scope for improvement. Future research could explore the development of more sophisticated and context-sensitive prompting methods. For instance, integrating persona prompting, where the model responds from the perspective of a specific persona or role, could potentially improve response relevance and alignment with human coders. Moreover, although we consciously refrained from providing few-shot examples in our study to test the model's inherent knowledge, incorporating few-shot prompting in future research could significantly enhance the model's performance and its alignment with human coders. Few-shot prompting involves providing the model with a few examples of the desired task, which can help the model better understand and adapt to nuanced instructions, which cannot be effectively communicated through specific instructions alone. Further refinement and testing of these prompting techniques could help in understanding their broader applicability and limitations.

To support ongoing and future research, it is essential to develop and refine analytical tools and workflows tailored to LLMs, making advanced text analysis more accessible and efficient for a broader range of researchers. The simplest step in the development of these models is the use of prompts. Prompts allow LLMs to generate responses based on pre-defined inputs, making it easy to guide the model's output without additional training. This approach is the most straightforward and requires minimal expertise, serving as a foundational technique for more complex methods. As mentioned: The next step could be the use of few-shot learning. Few-shot learning enables LLMs to efficiently learn new tasks and subject areas with only a few examples. This method is relatively straightforward to implement and requires moderate expertise. The next, more sophisticated step involves integrating knowledge bases in the form of Retrieval-Augmented Generation (RAG) systems. RAG systems provide real-time access to relevant information, thereby enhancing contextual understanding and classification accuracy. They significantly extend the capabilities of models by allowing access to a broader knowledge base, which is particularly beneficial for handling complex or specialised topics. Implementing RAG systems requires a higher level of expertise and is more resource-intensive than few-shot learning. Finally, fine-tuning the LLMs can further optimise their performance. By fine-tuning LLMs on specific data and use cases, the models can be adapted to the unique requirements and nuances of particular applications. This process could reduce misclassifications and increase the precision and reliability of the analyses substantially. But fine-tuning is the most demanding in terms of both effort and expertise, necessitating a deep understanding of the models and significant computational resources.

The progressive combination of prompts, few-shot learning, RAG systems, and fine-tuning can maximise the capability and flexibility of LLMs. However, it is crucial to find a sweet spot between accessibility and ease of use on one hand, and performance and complexity on the other. Each method, from simple prompts to complex fine-tuning, involves increasing levels of complexity and expertise, requiring careful consideration to balance these factors effectively.

Future research should also focus on developing techniques that ensure consistent and reliable outputs, especially when dealing with sensitive topics. Implementing strategies to standardise responses in the presence of offensive language or mentions of mental health issues is crucial, particularly for user-generated text that can employ various forms and styles of communication. Addressing issues related to repetitive characters, slang, and sensitive content will be essential in improving the consistency and reliability of LLM outputs.

With continuous advancements in LLMs and updated models being released in very short timespans, future work should focus on leveraging these enhanced capabilities for more specialised applications. Utilising the latest versions of LLMs, which likely come with improved contextual understanding and generation, can significantly enhance the accuracy and relevance of text analysis. This approach would ensure that researchers can fully exploit state-of-the-art model developments to achieve more precise analytical outcomes.

Additionally, fostering a collaborative research environment through sharing prompts, analysis scripts, datasets, and knowledge bases can significantly advance the field. By building on each other's work, incorporating RAG systems, or even fine-tuned models, researchers can develop more effective methods, thereby enhancing the overall performance and utility of LLMs in text classification tasks in various research fields. This collaborative approach will contribute to collective knowledge and progress in the domain, ensuring that advancements are shared and utilised to their fullest potential.

# APPENDICES

## ONLINE RESOURCES

Krause, T. (2024, August 12). Delving into Youth Perspectives on In-game Gambling-like Elements: A Proof-of-Concept Study Utilising Large Language Models for Analysing User-Generated Text Data. Retrieved from osf.io/gpy8s

## ADDITIONAL PLOTS

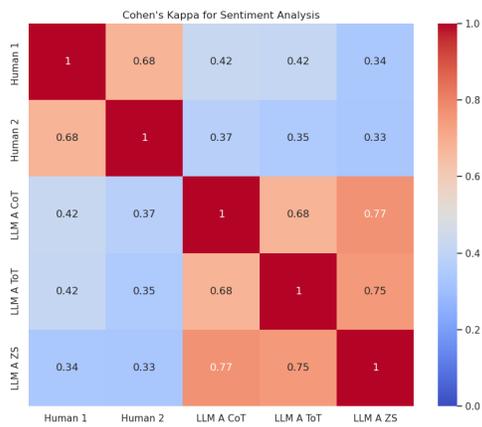
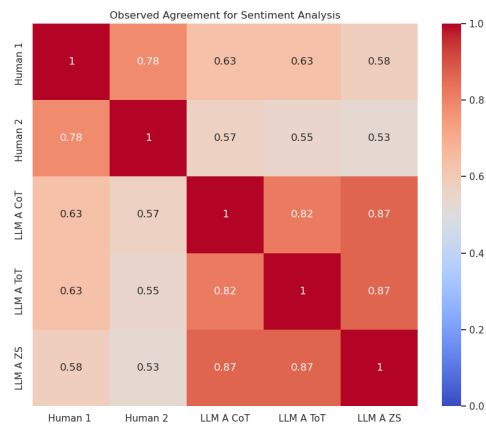
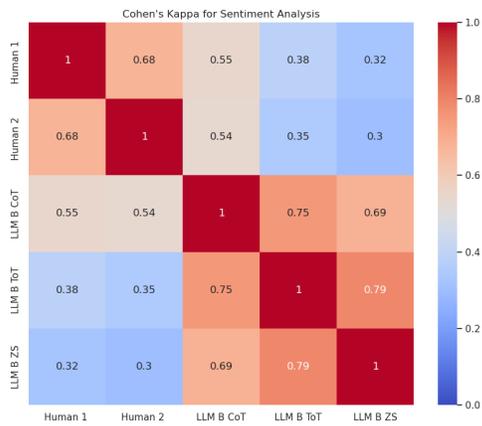
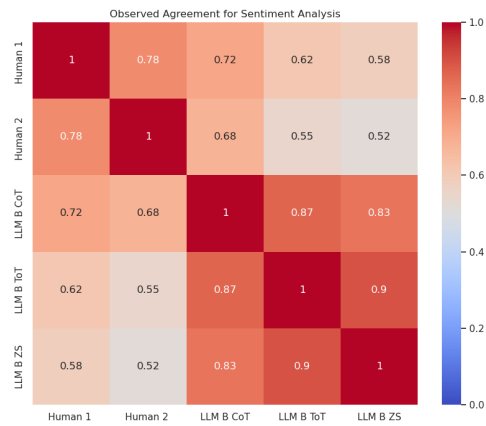
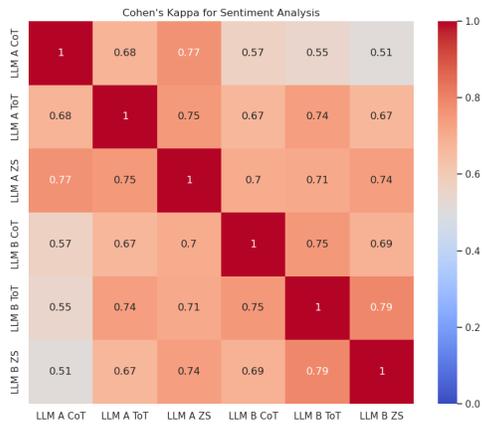
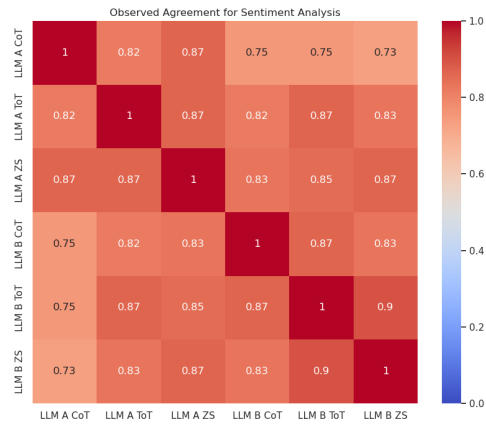

**Figure 7: Cohen's Kappa and Obs. Agreement between LLM Coding and Human Coding for ABSA (Test Data)**

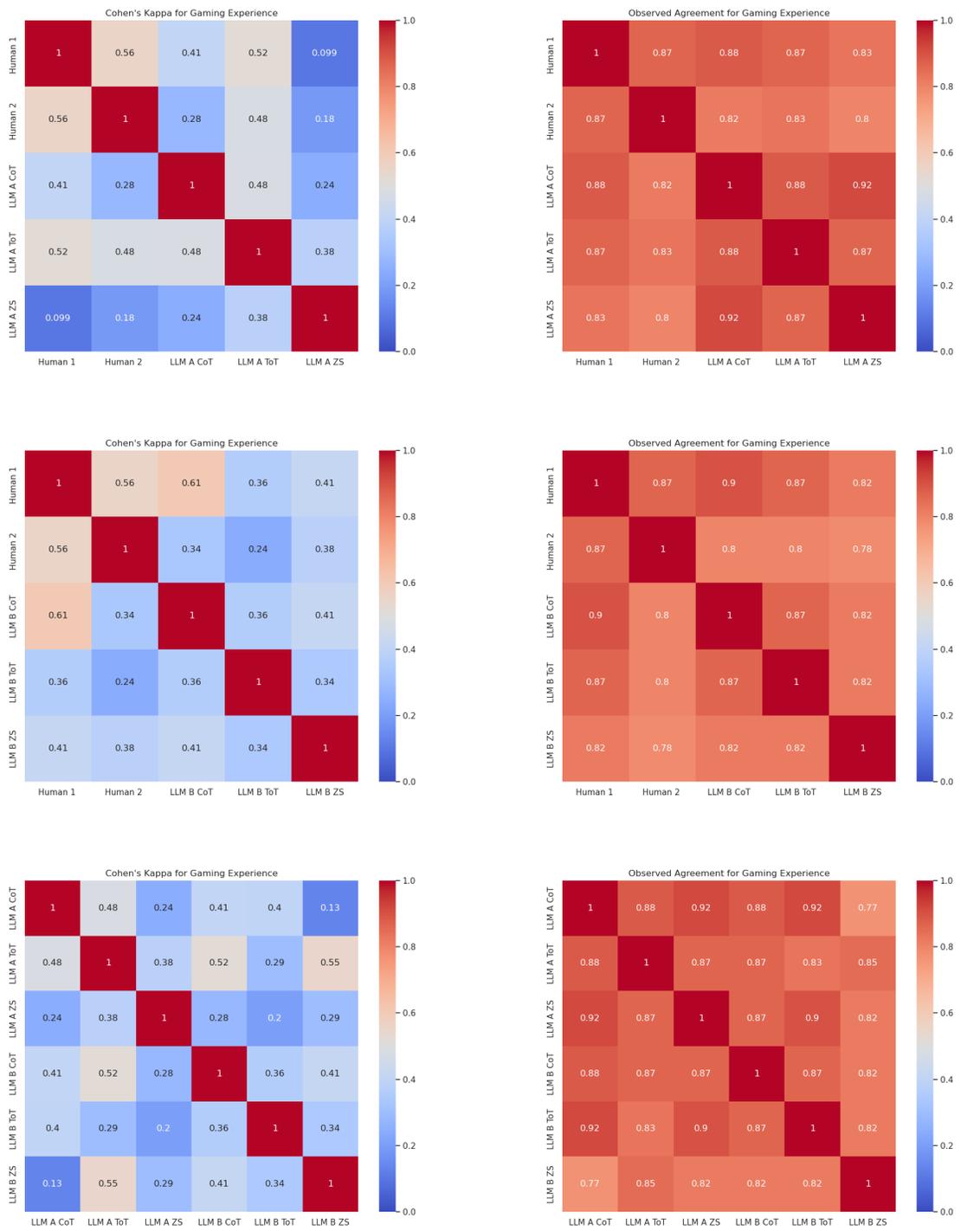

**Figure 8: Cohen's Kappa and Obs. Agreement between LLM Coding and Human Coding for Gaming Experience (Test Data)**

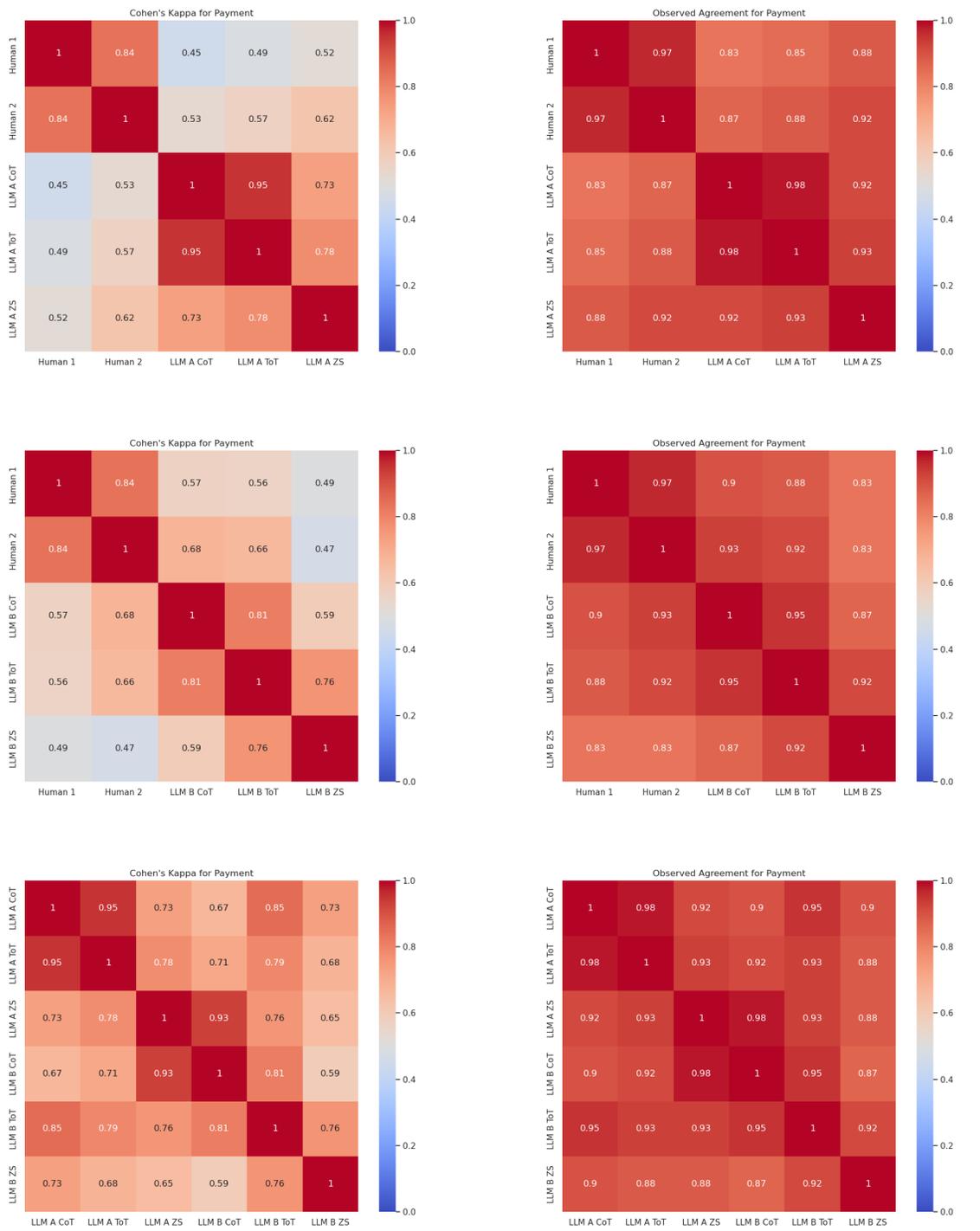

**Figure 9: Cohen's Kappa and Obs. Agreement between LLM Coding and Human Coding for Financial Engagement (Test Data)**

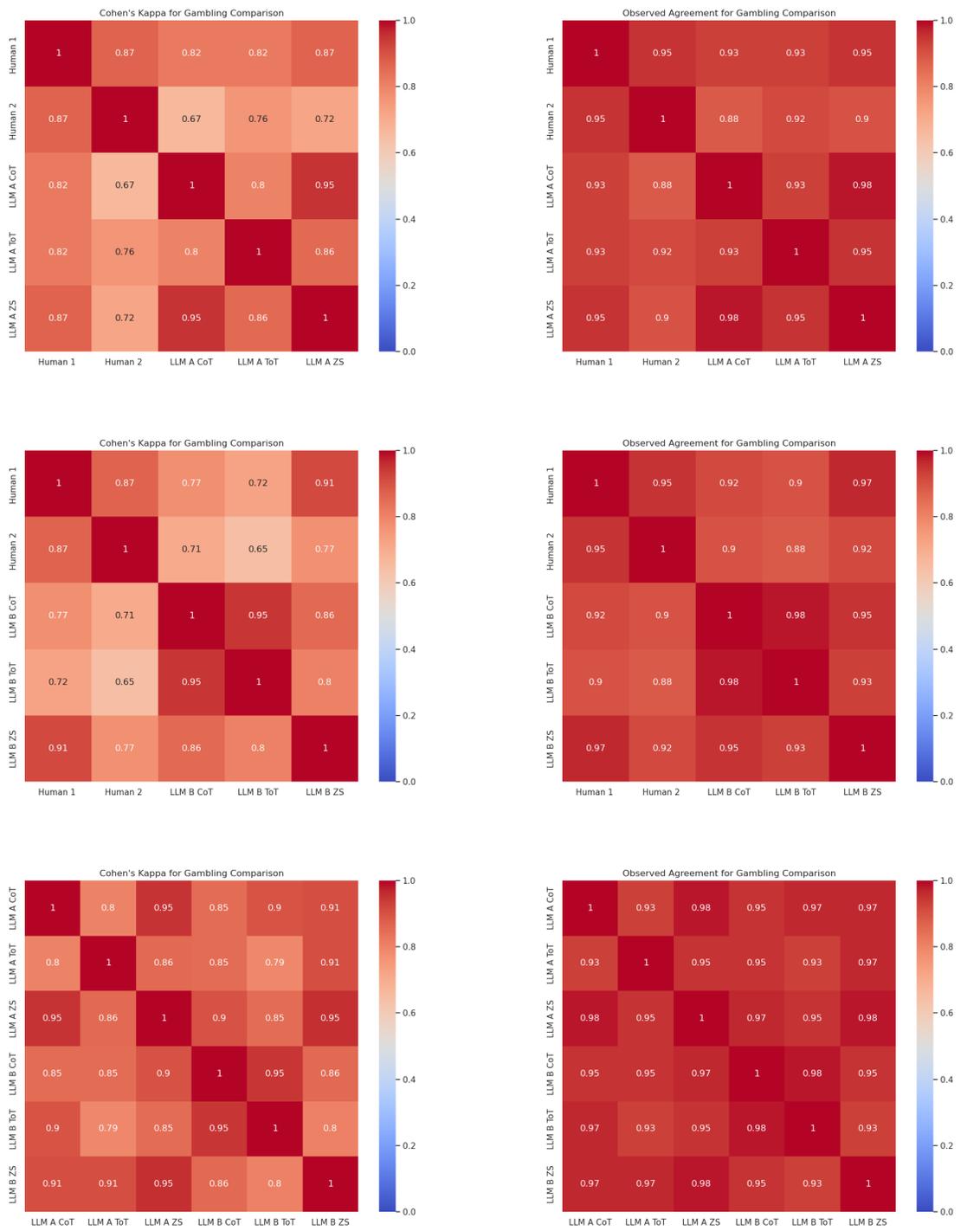

**Figure 10: Cohen's Kappa and Obs. Agreement between LLM Coding and Human Coding for Gambling Comparison (Test Data)**

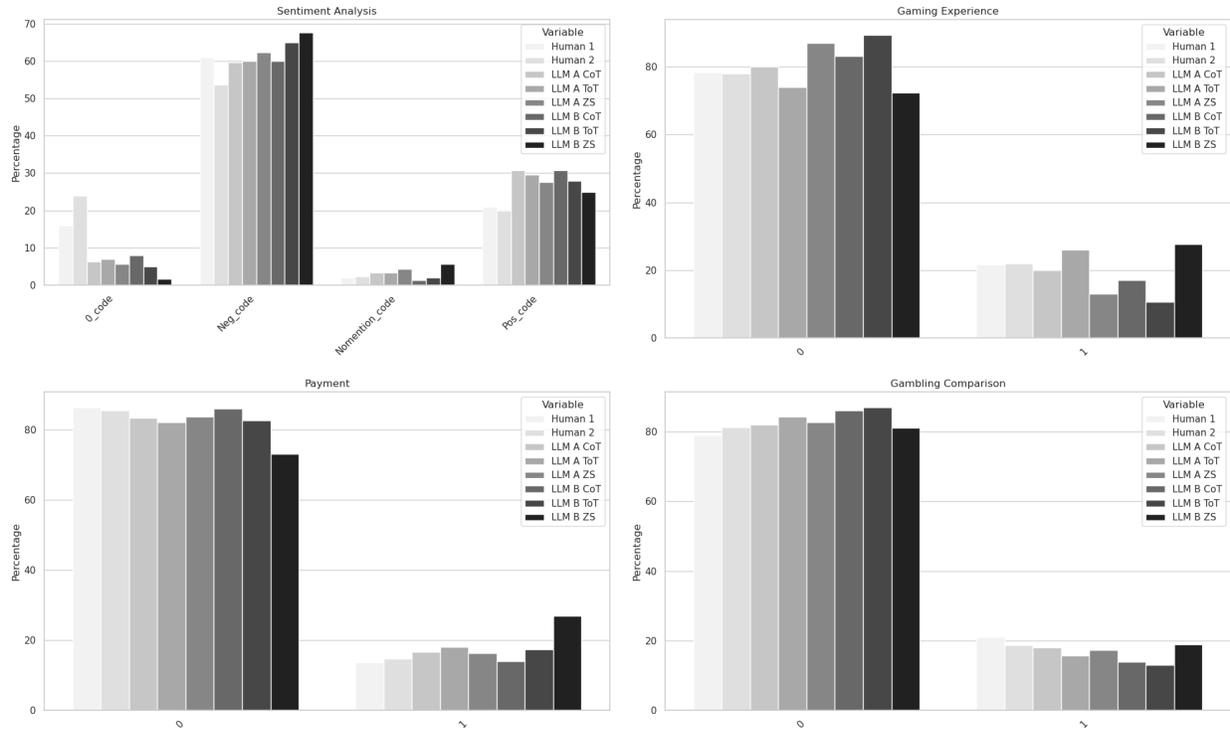

Figure 11: Distribution of Categories in the Training Set

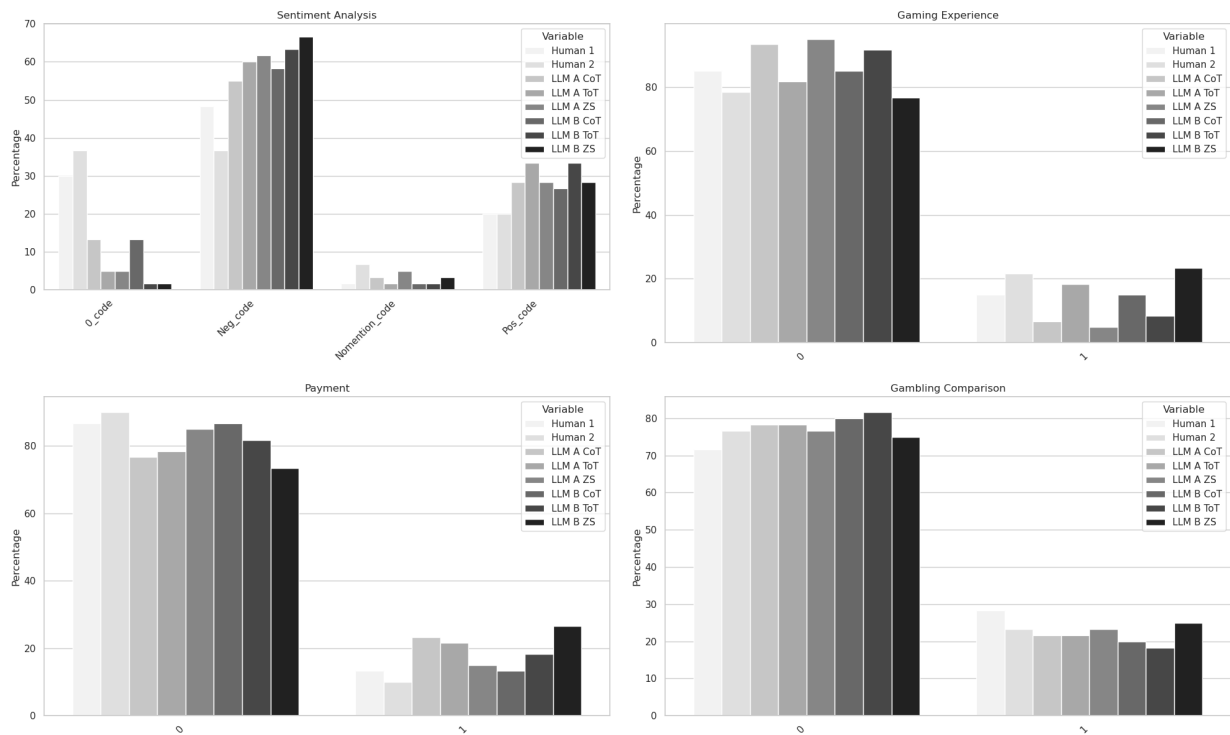

Figure 12: Distribution of Categories in the Test Set

# FINAL MODULAR PROMPTS TEMPLATES

## ABSA PROMPT

As part of a scientific study, your task is to analyze user text for sentiment
specifically for the concepts of lootboxes, card packs, loot crates, loot cases, Gacha, Mystery Boxes, or similar concepts.
Mentions of gambling should also be understood as references to these concepts.

For each text, determine the sentiment as positive, negative, or neutral exclusively towards these concepts.
Ensure that the analysis only pertain to these concepts and not to
- the entire monetization model,
- the game as a whole, or
- any other aspects of the game.

- If the text mentions any of these concepts with a positive sentiment towards them,
including expressions of disappointment or frustration about their absence, removal or the inability to access them,
provide a summary focusing on the positive aspects expressed and
end your response with 'Pos_code'.

- If the text mentions any of these concepts with a negative sentiment towards them,
including satisfaction or relief about their absence, removal or the inability to access them,
provide a summary focusing on the negative sentiments expressed and
end your response with 'Neg_code'.

- If the text mentions any of these concepts with a neutral, unclear, mixed, or no apparent sentiment,
provide a summary of the relevant aspects and end your response with '0_code'.

- If the text does not mention lootboxes, card packs, loot crates, loot cases, Gacha, Mystery Boxes, or similar concepts,
end your response with 'Nomention_code'.

Your final response should look like this:
Code: [ ]

## GAMING EXPERIENCE PROMPT

In this scientific analysis, your task is to determine
if the text mentions that concepts like lootboxes, card packs, loot crates, loot cases, Gacha, Mystery Boxes, or similar concepts
influence or do not influence the gaming experience.
The gaming experience can encompass game aspects such as overall quality, gameplay, game balance, competitive fairness or immersion.
The analysis should not treat the presence, enjoyment or use of lootboxes themselves as the gaming experience
but rather focus on whether it is mentioned that these mechanisms affect or do not affect the gaming experience of other aspects of the game.

Determine if a text explicitly discusses whether lootboxes or similar mechanisms
have an impact (or state that they do not have an impact) on the gaming experience.
This includes mentions of the effect on
quality, gameplay, game balance, competitive fairness, immersion or other aspects of the game.

For each text, you are to identify whether it mentions this aspect with a simple binary coding:
- If the text discusses the impact of these mechanisms on the gaming experience, assign a '1' to 'Gaming_Exp_Mention'.
- If the text does not mention the impact on the gaming experience, assign a '0' to 'Gaming_Exp_Mention'.

Your final response should look like this:

Code:

Gaming_Exp_Mention: [ ]

---

## FINANCIAL ENGANGEMENT PROMPT

In this scientific analysis, your task is to assess if user text directly mentions that the user has personally spent money on lootboxes, card packs, loot crates, loot cases, Gacha, Mystery Boxes, or similar mechanisms.

Implicit indications or merely suggested personal spending of money should not be counted.

It is important to note that mere interaction or usage of lootboxes without the explicit mention of spending money should not be counted.

For example, if a user talks about opening free lootboxes or using lootboxes obtained through gameplay without mentioning any personal financial transaction, it should not be counted.

For each text, determine whether it discusses the user's own financial expenditure on these elements.

- For texts that specifically mention the user personally spending money on lootboxes or similar mechanisms, assign a '1' under 'Payment_Willingness_Mention'.

- If there is no mention of the user personally spending money on lootboxes or similar mechanisms, assign a '0'.

Your final response should look like this:

Code:

Payment_Willingness_Mention: [ ]

---

## GAMBLING COMPARISON PROMPT

In this scientific analysis, your task is to strictly identify only explicit, direct comparisons or clear analogies to traditional gambling in relation to lootboxes, card packs, loot crates, loot cases, Gacha, Mystery Boxes, or similar mechanisms.

Ensure that:

- Subtle, implicit, or merely suggested comparisons are explicitly excluded.

- The inherent randomness of lootboxes and similar mechanisms, as well as any disappointment associated with their outcomes, should not be interpreted as similarities to or comparisons with gambling.

Focus on determining whether the text makes any unmistakable direct comparisons or analogies between gambling and these mechanisms.

This includes:

- Specific references to slot machines or casinos.

- Explicit mentions of addiction or gambling in the context of lootboxes or similar mechanisms.

- Discussion of legal or ethical issues directly comparing these mechanisms to gambling.

- When the term 'gambling' is used explicitly in context with these gaming mechanisms, it qualifies as a direct comparison, even if lootboxes or equivalent mechanisms aren't directly mentioned.

For each text reviewed, use the following binary coding to indicate the presence of these explicit connections:

- Assign a '1' under 'Gambling_Mention' for texts that directly make this connection.

- Assign a '0' if the text does not make this connection.

Your final response should look like this:

Code:

Gambling_Mention: [ ]

## COT PROMPT ADDITIONS

```
base_cot_prompt_a = """Let's think step by step. Explain your reasoning process carefully before you give your answer.
"""
base_cot_prompt_b = """
###
Remember to think step by step.

# ToT Prompt Additions
Imagine three different experts are collaborating to solve this task.
Each expert will write down one step of their thought process and share it with the group.
After sharing, all experts will proceed to the next step. This process will repeat until the task is complete.
At each step, every expert will assign a likelihood to their current assertion being correct.
If any expert realizes they are incorrect at any point, they will leave the group.
Once all experts have provided their analyses, you will review all three and
either present the consensus solution or your best guess.
Be sure to include a final answer code at the end.

The task is:
```

## TOT PROMPT ADDITIONS FOR LLAMA

```
Proceed to provide the complete analysis without stopping.

You must give a final answer code at the end of your answer.
```

## USER-TEXT INPUT PROMPT

```
###
Here is the text you need to analyze:
```{USERTEXT}```
```

## ACKNOWLEDGMENTS

### FUNDERS AND SUPPORTERS

We express our gratitude to the Bristol Hub for Gambling Harms Research for their financial support through the Research Innovation Fund Seedcorn Award 2023/24. This funding has been crucial for our project, "Delving into Youth Perspectives on In-game Gambling-like Elements."